\documentclass[10pt,journal,compsoc]{IEEEtran}                

\ifCLASSINFOpdf
\usepackage[pdftex]{graphicx}
\else
\fi
\usepackage{enumitem}
\usepackage{amsmath}
\usepackage{amsfonts}
\usepackage{caption}
\usepackage{hyperref}
\usepackage{url}

\hypersetup{
	colorlinks=true,
	citecolor=black,
	linkcolor=black,
	filecolor=black,      
	urlcolor=black,
}

\begin{document}

\title{ActFloor-GAN: Activity-Guided Adversarial Networks for Human-Centric Floorplan Design}
\author{
Shidong~Wang, Wei~Zeng,~\IEEEmembership{Member, IEEE}, Xi~Chen, Yu~Ye,\\ Yu~Qiao,~\IEEEmembership{Senior Member, IEEE}, and Chi-Wing~Fu,~\IEEEmembership{Member, IEEE}
\IEEEcompsocitemizethanks{

\IEEEcompsocthanksitem
S. Wang, X. Chen, and Y. Qiao are with Shenzhen Institute of Advanced Technology, Chinese Academy of Sciences. 
S. Wang is also with Shandong University. X. Chen is also with University of Chinese Academy of Sciences.
e-mail: sdwang96@gmail.com, \{xi.chen2, yu.qiao\}@siat.ac.cn.

\IEEEcompsocthanksitem
W. Zeng (corresponding author) is with The Hong Kong University of Science and Technology. email: weizeng@ust.hk.

\IEEEcompsocthanksitem
Y. Ye is with Tongji University. email: yye@tongji.edu.cn.

\IEEEcompsocthanksitem
C.-W. Fu is with the Chinese University of Hong Kong. email: cwfu@cse.cuhk.edu.hk.

}
\thanks{Manuscript received xx xx, 202x; revised xx xx, 202x.}}

\markboth{IEEE Transactions on Visualization and Computer Graphics}%
{Wang \MakeLowercase{\textit{et al.}}: Computational Design}
	
\definecolor{green}{rgb}{0.25,0.5,0.35}
\definecolor{purple}{rgb}{0.5,0,0.35}
\definecolor{red}{rgb}{0.9,0,0}
\definecolor{mred}{rgb}{0.5,0,0}
\definecolor{blue}{rgb}{0,0,0.9}
\definecolor{almond}{rgb}{0.94, 0.97, 0.97}
\definecolor{golden}{RGB}{240,232,225}
\definecolor{goldenbox}{RGB}{218,212,150}
\definecolor{linen}{RGB}{250,240,230}
\definecolor{linenbox}{RGB}{220,210,200}
\definecolor{lavender}{RGB}{230,230,250}
\definecolor{lavenderbox}{RGB}{200,200,220}
\definecolor{azure}{RGB}{240,255,255}
\definecolor{azurebox}{RGB}{210,225,225}
\definecolor{honeydew}{RGB}{240,255,240}
\definecolor{honeydewbox}{RGB}{210,225,210}

\newcommand{\zw}[1]{{\color{red}{ZW: #1}}}
\newcommand{\sd}[1]{{\color{red}{SD: #1}}}
\newcommand{\tvcg}[1]{{\color{black}{#1}}}
\newcommand{\TVCG}[1]{{\color{black}{#1}}}
\newcommand{\note}[1]{{\color{red}{#1}}}
\newcommand{\major}[1]{{\color{black}{#1}}}

\newcommand{\eg}{{\emph{e.g.}, }}
\newcommand{\ie}{{\emph{i.e.}, }}
\newif\ifnotes
\notestrue

\let\origcite\cite
\renewcommand{\cite}[1]{\ifnotes\mbox{\origcite{#1}}\else \origcite{#1}\fi}
\newcommand{\lorem}[1]{\ifnotes{\color{lgrey}{\texorpdfstring{#1}{#1}}}\fi}
\newcommand{\strike}[1]{\ifnotes{\color{mred}{\texorpdfstring{\sout{#1}}{#1}}}\fi}
\newcommand{\strikeg}[1]{\ifnotes{\color{grey}{\texorpdfstring{\sout{#1}}{#1}}}\fi}
\newcommand{\add}[1]{\ifnotes{\leavevmode\color{mred}{#1}}\else{#1}\fi}
\newcommand{\replace}[2]{\ifnotes{\strikeg{#1}\add{#2}}\else{#2}\fi}

\IEEEtitleabstractindextext{
\begin{abstract}
We present a novel two-stage approach for automated floorplan design in residential buildings with a given exterior wall boundary.
Our approach has the unique advantage of being human-centric, that is, the generated floorplans can be geometrically plausible, as well as topologically reasonable to enhance resident interaction with the environment.
From the input boundary, we first synthesize a human-activity map that reflects both the spatial configuration and human-environment interaction in an architectural space.
We propose to produce the human-activity map either automatically by a pre-trained generative adversarial network (GAN) model, or semi-automatically by synthesizing it with user manipulation of the furniture.
Second, we feed the human-activity map into our deep framework \emph{ActFloor-GAN} to guide a pixel-wise prediction of room types.
We adopt a re-formulated cycle-consistency constraint in \emph{ActFloor-GAN} to maximize the overall prediction performance, so that we can produce high-quality room layouts that are readily convertible to vectorized floorplans.
Experimental results show several benefits of our approach.
First, a quantitative comparison with \major{prior methods} shows superior performance of leveraging the human-activity map in predicting piecewise room types.
Second, a subjective evaluation by architects shows that our results have compelling quality as professionally-designed floorplans and much better than those generated by existing methods in terms of the room layout topology.
Last, our approach allows manipulating the furniture placement, considers the human activities in the environment, and enables the incorporation of user-design preferences.
\end{abstract}

}
	
\maketitle
\begin{figure}[t]
	\centering
	\includegraphics[width=0.495\textwidth]{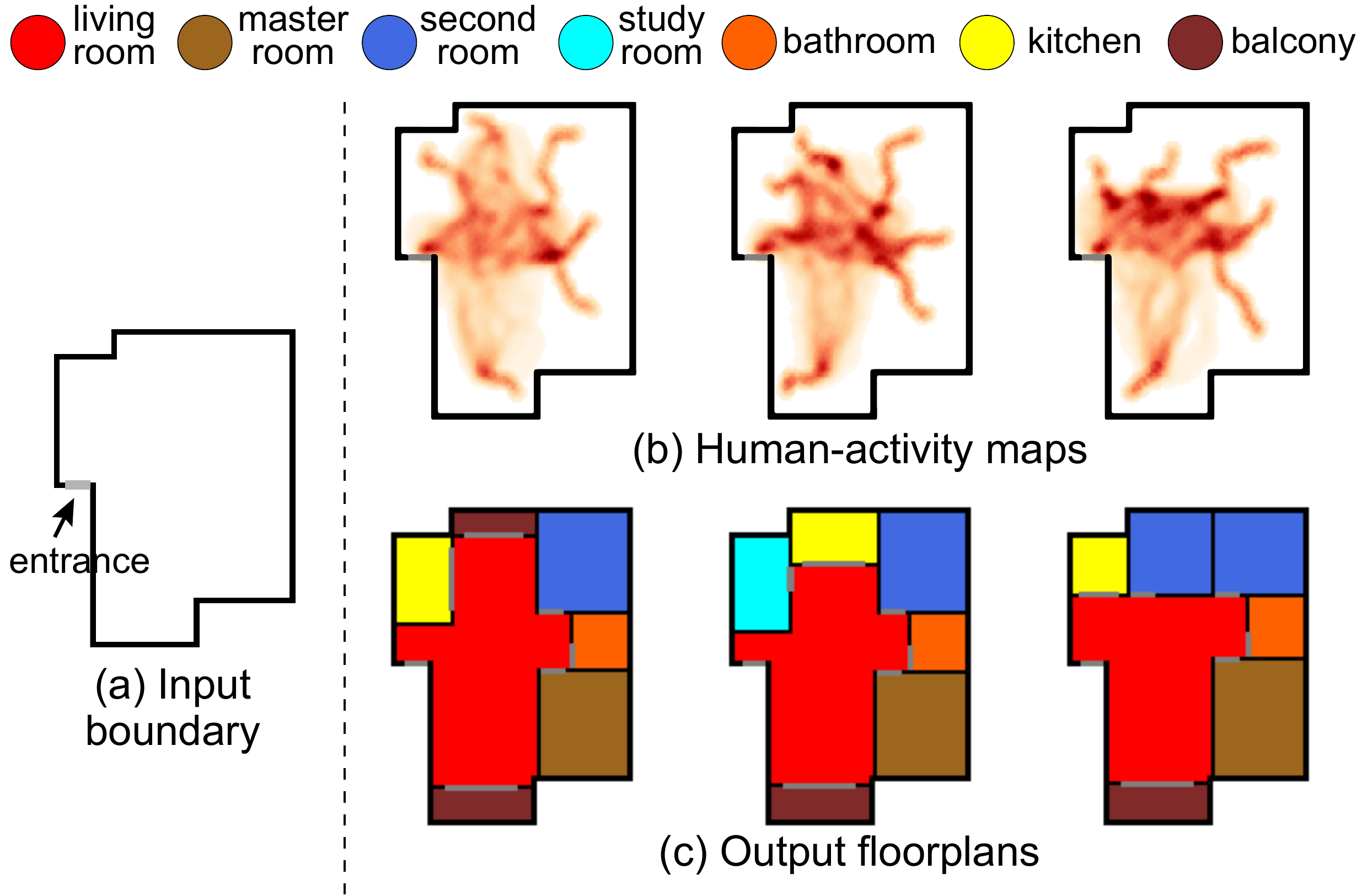} \\
	\vspace*{-2mm}
	\caption{Given an exterior wall boundary as input (a), our method can generate diverse floorplans (c) that are not only geometrically plausible but also topologically reasonable in correspondence with the human-activity maps (b).}
	\label{fig:aim}
	\vspace*{-4mm}
\end{figure}

\IEEEraisesectionheading{
\section{Introduction}}
\IEEEPARstart{C}{omputational} architecture design attempts to automate the generation of architectural layouts that define the spatial configuration of rooms, walls, and doors (Figure~\ref{fig:aim}(c)) within a given exterior wall boundary (Figure~\ref{fig:aim}(a)). 
Automated floorplan generation for residential buildings has been popular for decades~\cite{levin_1964_use}.
The topic has attracted researchers not only in architecture (\eg\cite{arvin_2002_modeling, michalek_2002_architectural, rodrigues_2013_approach}), but also in  related fields like computer graphics (\eg\cite{bao_2013_generating, merrell_2010_computer, wu_2018_miqp}).
In general, a two-stage approach is adopted: first collect a set of constraints related to the geometric (\eg room sizes and positions) and/or topological (\eg room adjacency) properties of the room layout~\cite{michalek_2002_architectural}, then optimize to produce a room layout that fulfills the constraints.

Conventional methods exploit various design constraints and optimization techniques.
However, there still remain many challenges.
First, floorplan design tends to be ill-defined and over-constrained~\cite{arvin_2002_modeling}.
Common geometric and topological properties can be modeled relatively easily by means of objectives such as cost and performance.
However, architectural designers \major{are also concerned with} the aesthetic and usability aspects of a room layout, which are generally more difficult to describe formally~\cite{michalek_2002_architectural}.
Second, rooms and walls in a floorplan are deformable shapes that often do not have pre-defined dimensions, thus bringing additional difficulty for the subsequent optimization model to converge.
Last but not least, the design space of floorplans is diverse$-$there exist a wide range of feasible room layouts for the same building boundary.
Taking Figure~\ref{fig:aim}(a) as an example, there exist multiple feasible floorplans (Figure~\ref{fig:aim}(c)).
\major{Architectural designers} often want to explore diverse design options rather than having only one single solution.

Recently, deep learning methods are being increasingly used for automated floorplan design.
On the one hand, vast amounts of floorplans in vector-graphics format have been manually curated (\eg CubiCasa5K~\cite{kalervo_2019_cuicasa5k} and RPLAN dataset~\cite{wu_2019_data-driven}) or automatically mined from raster images (\eg \cite{liu_2017_vector, zeng_2020_deep}), providing reliable data for neural networks to learn.
The floorplans are from real-world designs, which reflect good practices of room layout design by professional designers.
On the other hand, various efforts have been put on modeling room layouts into network consumable formats, and on designing neural networks to automate floorplan generation.
For example, Wu et al.~\cite{wu_2019_data-driven} developed an iterative prediction model to determine the room connections and positions, and an encoder-decoder network to compute the room sizes and wall locations.
Later, House-GAN~\cite{nauata_2020_house-gan} and Graph2Plan~\cite{hu_2020_Graph2Plan} explored generative networks to learn room relations. 
However, these deep-learning-based methods typically involve network structures that are complicated.
\tvcg{Taking the work of Wu et al.~\cite{wu_2019_data-driven} as an example, their model consists of four deep networks, including a regression network for locating the living room, another regression network for predicting the other room positions and types, a continuing network to determine if the process should continue, and an encoder-decoder network to predict walls}.
Moreover, architectural designers often want to explore various design options, whilst many existing methods fail to meet the requirements.

\major{Modern architecture is an exploratory process of creating designs for people~\cite{history_architecture}.
Human activities and their behavior preferences are always key concerns at the preliminary phase of the design process~\cite{hu_2020_predicting}.
Following the protocol of architecture design,} we propose a human-centric approach \major{using human-activity maps, as guidance for floorplan design.}
Being human-centric, a floorplan should be geometrically plausible and satisfy topological requirements, such that the residents can have pleasant interactions with the environment~\cite{arvin_2002_modeling}.
As suggested by a collaborating architect, we leverage the concept of human-activity map (see Figure~\ref{fig:aim}(b)) that describes human spatial behavior in an architectural space~\cite{franz_2008_from}, reflecting both the spatial configuration and human-environment interaction of room layouts.
Given an input boundary, we first provide two alternative approaches for generating human-activity maps:
(i) a fully-automated method via a generative neural network, and (ii) a semi-automatic method via a user interface.

\tvcg{Next, we present a two-stage approach that produces vectorized floorplans from building boundaries and human-activity maps.
At the first stage, we design a new deep framework, namely \emph{ActFloor-GAN}, that takes a human-activity map to guide the floorplan generation from the input boundary.
All inputs and outputs of \emph{ActFloor-GAN} are raster images, allowing us to adopt a re-formulated cycle-consistency constraint, including (i) an adversarial loss to encourage the generated floorplans to be indistinguishable from real floorplans; (ii) a cycle-consistency loss to enforce the forward-backward consistency when translating between the boundary and floorplan domains; and (iii) an identity loss to regularize the generator to form a near identity mapping when a real sample of the target domain is provided as an input to the generator.
We train \emph{ActFloor-GAN} on the RPLAN dataset~\cite{wu_2019_data-driven} and show that the model improves the overall performance in terms of the pixel-wise prediction accuracy compared with \major{prior methods}.
At the second stage, we convert the pixel-wise predictions from \emph{ActFloor-GAN} into vectorized floorplans for convenient usage by architects.
Subjective evaluation by architects shows that the floorplans by our approach have compelling quality as the professionally-designed floorplans, and are more preferred than those by Wu et al.~\cite{wu_2019_data-driven}.}

The main contributions of our work are as follows:

\begin{itemize}
\item
We design a novel deep learning method that employs the human-activity map to guide the floorplan design (Sec.~\ref{sec:floorplan_generation}).
To the best of our knowledge, this is the first attempt that considers human activities in computational floorplan design, in which we look at both the spatial configuration and human-environment interaction.

\vspace{1.5mm}
\item
We develop automatic and semi-automatic approaches to synthesizing the human-activity map (Sec.~\ref{sec:activity_map}).
Experimental results show that the automatic approach produces accurate piece-wise room predictions in comparison to \major{prior methods} without guidance by human-activity maps and adoption of cycle-consistency constraints (Sec.~\ref{ssec:quantitative}), and generates vectorized floorplans of high quality that are comparable to professional designs, as evaluated by architects (Sec.~\ref{ssec:qualitative}).
Also, the semi-automatic approach enables flexibility to generate diverse human-centric floorplans (Sec.~\ref{ssec:diversity}).

\end{itemize}


\section{Related Work}
\label{sec:related_work}

\begin{figure*}[t]
\centering
\includegraphics[width=0.95\textwidth]{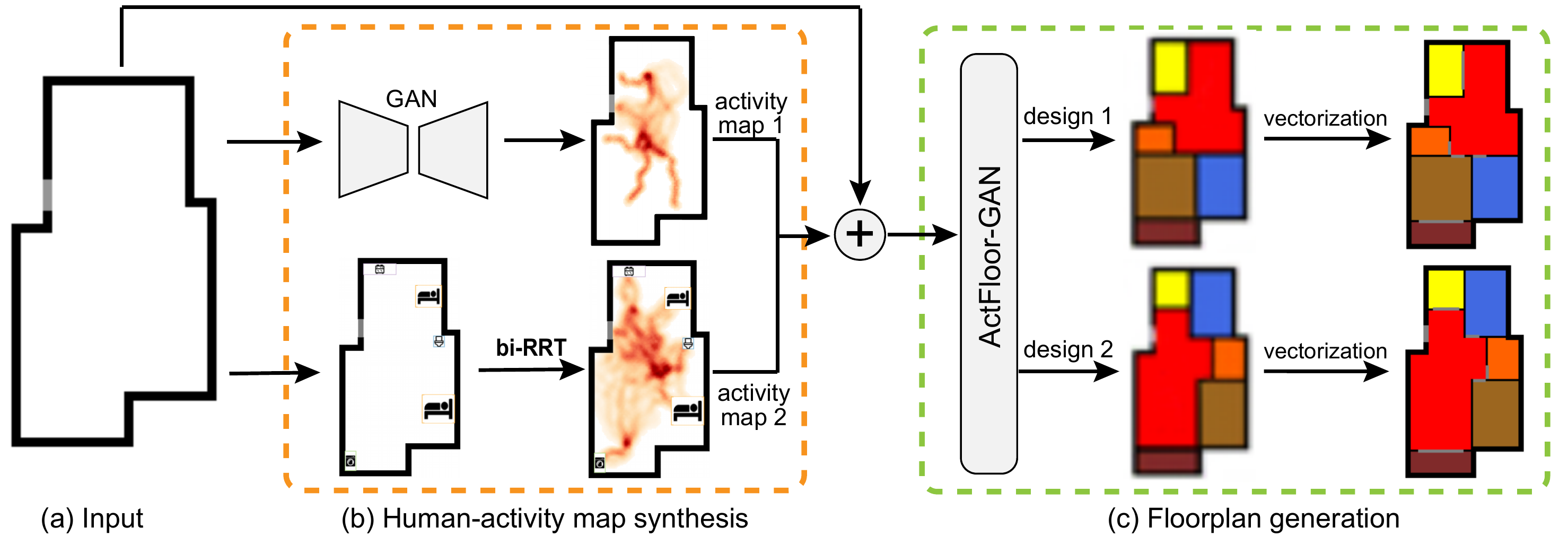} \\
\vspace*{-3mm}
\caption{
An overview of our human-centric floorplan design framework.
Given an input building boundary (a), we first synthesize the human-activity map (b) using either an end-to-end GAN model (b-top) or bi-RRT~\cite{lavalle_1998_rapidly-exploring} based on user manipulation of furniture placement (b-bottom).
Next, in the floorplan generation stage (c), our ActFloor-GAN consumes the building boundary \& human-activity map, and predicts the piecewise room segmentations that can be vectorized into the final floorplans.
ActFloor-GAN can produce diverse designs corresponding to different human-activity maps.
}
\vspace{-4mm}
\label{fig:workflow}
\end{figure*}

\noindent
\textbf{Computational Layout Design}.
Layout design is an important area of inquiry that aims to find a feasible spatial configuration for a set of interrelated objects.
Research on computational layout design spans a wide range of scenarios, such as user interface (\emph{e.g.},~\cite{swearngin_2018_rewire, lee_2020_guicomp}), visualization (\emph{e.g.},~\cite{bylinskii_2017_learning, chen_2021_composition}), urban planning (\emph{e.g.},~\cite{yang_2013_urban, peng_2016_computational}), indoor scene (\emph{e.g.},~\cite{wang_2018_deep, li_2019_grains}), \TVCG{document layout (\eg\cite{patil2020read}), and shape structures (\eg\cite{li2017grass})}.
In this work, we focus on the computational architecture layout design, specifically to generate room layouts for an input building boundary.

The design space of floorplan can be decomposed into two aspects: \emph{geometry} that constrains the position and size of each room, and \emph{topology} that concerns the logical relationships between rooms~\cite{michalek_2002_architectural}.
Conventional methods for computational architecture layout design typically adopt a two-stage approach---first determine the relevant constraints, then optimize the layout.
Merrell et al.~\cite{merrell_2010_computer} sampled the topological relationship of room adjacencies from real-world data, then applied the relationship to optimize room layouts via stochastic optimization.
Bao et al.~\cite{bao_2013_generating} characterized local shape variations as constrained optimization, and linked them to a portal graph to facilitate layout exploration.
Wu et al.~\cite{wu_2018_miqp} identified high-level constraints of room size, position, and adjacency, and formulated the constraints as a mixed-integer quadratic programming problem.

Though being automatic, the above approaches typically suffer from the limitations of oversimplified constraints and crudely estimated parameters~\cite{bao_2013_generating}.
Architectural design inherently involves subjective decisions such as aesthetics and domain expertise, involving a complex and high-dimensional solution space.
Data-driven approaches can suppress the deficiency by learning fine layouts from the existing databases and applying the learned knowledge to generate new layouts.
House-GAN~\cite{nauata_2020_house-gan} builds a relational generative adversarial network that takes a bubble diagram to encode the room adjacency as constraints and generates room layouts that satisfy the constraints.
The high-level constraints are extracted from a linguistic expression that describes the house details~\cite{chen_2020_intelligent}.
Both methods leave to users the selection of the generated boundaries, without specifying the boundary as a constraint that is commonly adopted by architects.
Graph2Plan~\cite{hu_2020_Graph2Plan} addresses the limitation by associating the boundaries using an adjacency layout graph and retrieving the associations from an existing database.
The method, however, requires a detailed structured representation, and the quality of the generated room layouts by the method can be severely affected by the retrieved boundary.
Also, the topology graph alone may not sufficiently capture the human spatial behavior and experience~\cite{franz_2008_from}.

Alternatively, Wu et al.~\cite{wu_2019_data-driven} directly took a building boundary as input without requiring manual preparation of the topology graph and high-level room adjacency constraints. 
The method employs an iterative prediction model that utilizes a CNN to first regress the living room center, an encoder-decoder network to iteratively determine the center of each room, and another CNN to control the iterative process.
However, the iterative process has a long-running time, as it typically requires many iterations to cease.
In this work, we propose a new generative adversarial network (GAN) model that directly predicts piecewise room layouts.
More importantly, our model encapsulates the human-activity maps to constrain the occupants' interactions with the environment and render more realistic room layouts.
We show the superior performance of our approach over Wu et al.~\cite{wu_2019_data-driven} in terms of preference by five architects.

\vspace{1.5mm}
\noindent
\textbf{Human-Activity Map}.
\tvcg{In a typical architecture design process, architects first predict functions of interior space to estimate potential human activities, based on which they draw initial bubble charts that reflect the functional arrangement of the space, and finally generate detailed floorplans~\cite{Hertzberger_1991_lessons}.
That is, human activities are considered at the preliminary phase of the design process for interior architecture design~\cite{hu_2020_predicting}, even prior to topology as `form follows functions'~\cite{corbusier_1927_vers}.
Many studies leverage bubble charts as guidance for floorplan design (\eg\cite{merrell_2010_computer, hu_2020_Graph2Plan, nauata_2020_house-gan}), yet few attentions have been paid to human activities.}
This work exploits human-activity map that describes the spatial behavior of residents in an architectural space, and encodes the relationship between the environment (room layouts and furniture locations) and the residents' activities~\cite{franz_2008_from}.
\major{To a certain extent, human-activity maps reveal users’ behavior intensity in buildings, helping to locate frequently-used walking paths and dwell points. The information is important concerns in architectural design process~\cite{kim1999spatial}.}
Some studies have employed the human-activity map to guide architectural design. 
For example, Feng et al.~\cite{feng_2016_crowd-driven} optimized the human crowd properties of mobility, accessibility, and coziness when designing mid-scale layouts such as shopping malls.
However, few works have considered human activities in apartment-scale residential design, and a particular challenge is that human activity is hard-to-quantify~\cite{berseth_2019_interactive}.
Conventional methods for generating human-activity maps typically rely on time-consuming agent-based crowd simulations, such as~\cite{reynolds_1987_flocks, pelechano_2007_controlling}.
Recently, Qi et al.~\cite{qi_2018_human-centric} incorporated a human-activity probability map computed from the bi-directional rapidly-exploring random
tree (bi-RRT)~\cite{lavalle_1998_rapidly-exploring} to generate functional and natural residential rooms.
Hu et al.~\cite{hu_2020_predicting} designed a neural network to learn environment-crowd relationships from crowd simulations. 

This work adopts the human-activity map to guide architecture layout design.
Specifically, we offer two approaches to generate human-activity maps from an input boundary:
(i) an automatic method with a GAN model trained from synthetic human-activity maps;
and (ii) a semi-automatic method using bi-RRT based on user-specified furniture locations.
These alternatives enable the generation of diverse floorplan designs to match the architect's preference.

\vspace{1.5mm}
\noindent
\textbf{Generative Adversarial Networks}.
GANs~\cite{Goodfellow_2014_GANs} are a class of methods that train deep generative models based on the game-theoretic min-max principles.
Basically, the generator \emph{G} translates a sample from the generator distribution $P_G$ to data distribution $P_{data}$ and the discriminator \emph{D} determines if a sample belongs to $P_{G}$ or $P_{data}$.
Typically, we train $G$ and $D$ together, so that both neural networks can improve together.
In practice, GANs have been successfully applied in many fields, including image translation (\emph{e.g.},~\cite{Li_2016_Precomputed, isola_2016_Image-To-Image}) and image super-resolution (\emph{e.g.},~\cite{Ledig_2017_Photo-Realistic, karras_2018_progressive}).

Recently, GANs have also been employed to facilitate layout design, \emph{e.g.}, content-aware graphic design layout generation~\cite{Zheng_2019_Content-Aware}.
LayoutGAN~\cite{li_2020_layoutgan, li_2020_attribute} models the geographic relations of 2D elements and learns to arrange them in layout designs.
The method is, however, limited to simple 2D shapes of points, triangles, and rectangles, whilst floorplan design requires arranging more complex polygons within a constrained building boundary.
The most related approach to our work is ArchiGAN~\cite{Chaillou_2019_archigan}, which directly predicts pixel-wise room types using the \emph{pix2pix} framework~\cite{isola_2016_Image-To-Image}.
We experimented with the model on the RPLAN dataset~\cite{wu_2019_data-driven}.
However, its predicted room types are too noisy, so are infeasible for vectorization.
This is probably because the mapping from building boundary to realistic floorplan is one-to-many rather than one-to-one.
To address the problem, we re-formulate the cycle consistency constraints~\cite{zhu_2017_cyclegan} to improve the piece-wise room predictions and incorporate the human-activity maps in our \emph{ActFloor-GAN} model to guide the floorplan generation.
\section{Overview}
\label{sec:overview}

Figure~\ref{fig:workflow} overviews our framework for human-centric floorplan design.
It requires only a building boundary as input (Figure~\ref{fig:workflow}(a)), which is easier to prepare in comparison to high-level constraints, \emph{e.g.}, topology graph of room adjacency, needed in~\cite{nauata_2020_house-gan, chen_2020_intelligent,hu_2020_Graph2Plan}, which require domain knowledge from the architects.
\tvcg{This work employs the RPLAN dataset~\cite{wu_2019_data-driven}, which contains 80K floorplans from real-world residential buildings.
Each floorplan is represented as a four-channel 256$\times$256 raster image that encodes an \emph{inside mask}, a \emph{boundary mask}, a \emph{room category mask}, and a \emph{room IDs mask}.
The room types include the \emph{Living room, Master room, Second room, Study room, Bathroom, Kitchen, Balcony, and Other room}.
The dataset, however, does not include human activity information, which is crucial in our framework.
We opt to synthesize the human-activity maps based on the existing room layouts.}
As such, we propose a new two-stage approach with the following two major modules.

\begin{figure*}[ht]
\centering
\includegraphics[width=0.995\textwidth]{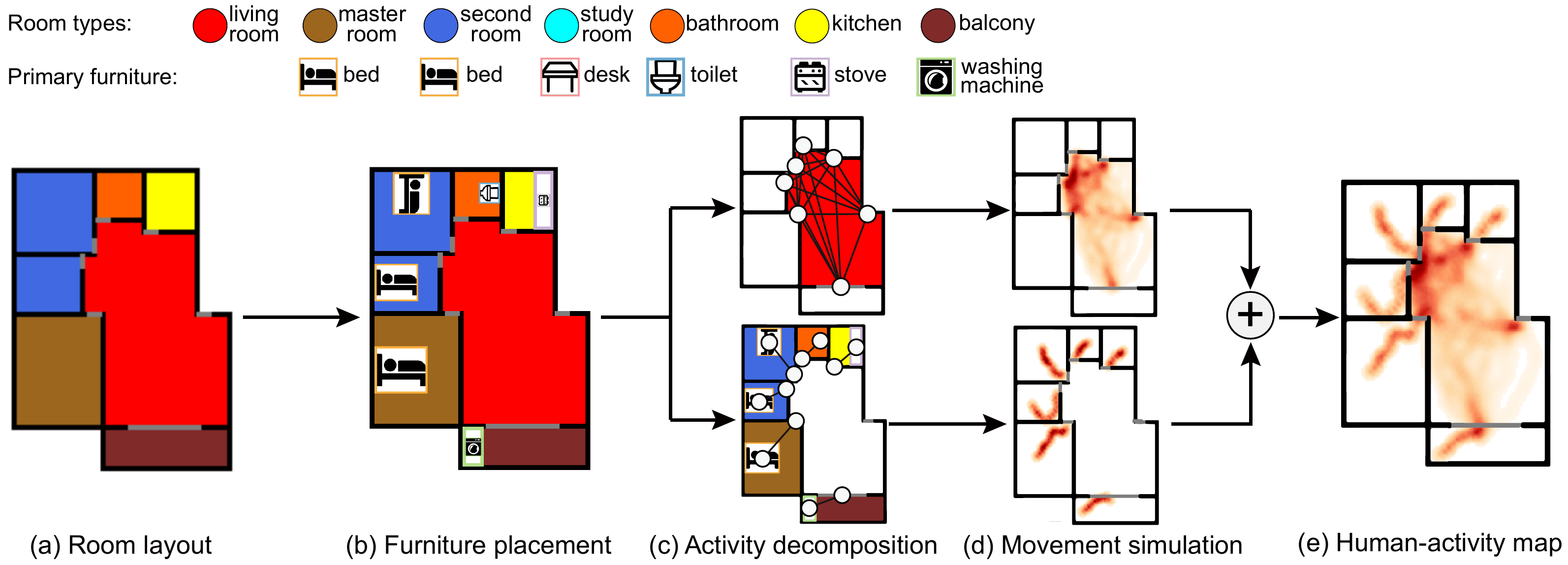} \\
\caption{The synthesis process of the human-activity map. Taking a floorplan from the RPLAN dataset~\cite{wu_2019_data-driven} as input (a), we first place primary furniture in each room (excluding the living room) according to conventional rules (b). Next, we decompose the human activities into two parts: in living room and in other rooms (c), and adopt bi-RRT~\cite{lavalle_1998_rapidly-exploring} to simulate the resident movements in rooms accordingly (d). Lastly, we combine the  simulated movements in different rooms together, yielding a human-activity map (e).
Note that the boundary and wall outline in (d) \& (e) are for reference only.}
\label{fig:activity_map}
\end{figure*}

\vspace{1mm}
\noindent
\emph{(i) Human-activity map synthesis} (Figure~\ref{fig:workflow}(b)).
Well-designed floorplans improve the interaction of the residents with the environment, such as increasing the visibility of an area from observation points~\cite{berseth_2019_interactive} and enabling efficient evacuation~\cite{hu_2020_predicting}.
This work considers human activities to guide the furniture placement to create functional and natural residential rooms~\cite{qi_2018_human-centric}.
We incorporate and develop both automatic and semi-automatic methods for generating human-activity maps.

\begin{itemize}
\item
\emph{Automatic}.
We train an end-to-end GAN model by pairing building boundaries from the real-world floorplan dataset RPLAN~\cite{wu_2019_data-driven} with corresponding human-activity maps synthesized using bi-RRT~\cite{lavalle_1998_rapidly-exploring}.
The synthetic human-activity maps consider room-room and room-furniture relations, where the furniture pieces are arranged based on common practices suggested by our collaborating architect.
The trained GAN model takes a building boundary as input, and automatically outputs a human-activity map (Figure~\ref{fig:workflow}(b-top)).

\vspace{1.5mm}
\item
\emph{Semi-automatic}.
We develop an interactive interface for users to arrange furniture, \emph{e.g.}, beds, toilets, and desks, etc., within the building boundary.
Our system will then synthesize the corresponding human-activity map using bi-RRT (Figure~\ref{fig:workflow}(b-bottom)), which can substitute the automatically-generated activity map (Figure~\ref{fig:workflow}(b-top)) and
enhance the flexibility of our approach for on-demand floorplan design.
\end{itemize}

\vspace{1mm}
\noindent
\emph{(ii) Floorplan generation} (Figure~\ref{fig:workflow}(c)).
Given the building boundary and human-activity maps as inputs, our goal is to automatically generate diverse floorplans that meet the conditions of human-environment interaction embedded in the activity map. 
The floorplan generation is a twofold process.
First, we train our \emph{ActFloor-GAN} model to predict pixel-wise room types within the boundary and take the human-activity map to guide the room-type prediction via our re-formulated cycle-consistency constraints.
Quantitative experimental results reveal the superior performance of our method in piecewise room-type prediction than \major{prior methods} without the human-activity map or the re-formulated cycle-consistency constraints. 
Next, we use a post-processing module to refine the piecewise room-type predictions into vectorized floorplans.
Subjective evaluation by architects shows that the floorplans produced by our approach are comparable with the professionally-designed ones, and better than those by Wu et al.~\cite{wu_2019_data-driven}.

\section{Human-Activity Map}
\label{sec:activity_map}

This section first introduces conventional rules adopted to synthesize the human-activity map from an existing floorplan (Sec.~\ref{ssec:activity_map}).
Next, we present how we generate the human-activity maps based on unseen building boundaries in the prediction stage (Sec.~\ref{ssec:activity_map_from_building_boundary}).

\subsection{Synthesis from Room Layout}
\label{ssec:activity_map}

We utilize the room layout information in the RPLAN dataset~\cite{wu_2019_data-driven} to synthesize human-activity maps.
As depicted in Figure~\ref{fig:activity_map}, the synthesis has two steps: furniture placement (Sec.~\ref{sssec:furniture_place}) and movement simulation (Sec.~\ref{sssec:movement_simulate}).

\subsubsection{Furniture Placement}
\label{sssec:furniture_place}

Human-activity map reflects the residents' interactions with the environment, specifically room layouts and furniture in the rooms.
However, the RPLAN dataset contains only the room layout information, but not the furniture locations.
To fill the gap, we place furniture in the rooms by carefully considering various requirements.
Typically, an apartment contains many different furniture, e.g., bed, and toilet.
Detailed furniture configuration can produce fine-grained resident movements.
However, excessive furniture in a room could over-constrain human activities, and in turn, affect the network's generalizability in the floorplan generation. 
We opt to make a trade-off between the movement details and model generalizability, by identifying and arranging primary furniture in each room except the living room.

In consulting our collaborating architect, we formulate the following furniture arrangement rules:

\begin{itemize}[leftmargin=*]
\item
\emph{Master \& Second rooms}: The primary furniture is a \emph{Bed}, of which the size is determined in correspondence with the room size and the position aligns with the wall on the opposite or diagonally opposite side of the room entrance.

\item
\emph{Study room}: The primary furniture is a \emph{Desk}, of which the size is determined in correspondence with the room size and the position is randomly allocated either beside or opposite to the room entrance.

\item
\emph{Bathroom}: The primary furniture is a \emph{Toilet}, of which the size is fixed and the position is typically allocated on the opposite or diagonally opposite side of the room entrance.

\item
\emph{Kitchen}: The primary furniture is a \emph{Stove}, of which the length equals to one of the wall lengths, and the position aligns with the wall opposite or diagonally opposite to the room entrance.

\item
\emph{Balcony}: The primary furniture is a \emph{Washing machine}, which has a fixed size and is positioned on one of the sides of the balcony.
\tvcg{In many Asian cities, from which most floorplans in the RPLAN dataset were collected, it is common to put a washing machine on the balcony.}

\end{itemize}

To enrich the diversity of the synthesized human-activity maps, we provide multiple optional locations for each piece of furniture in a room.
Figure~\ref{fig:activity_map}(b) shows an example of furniture placement.
Here, the room entrances are retrieved from the RPLAN dataset~\cite{wu_2019_data-driven}.

\subsubsection{Movement Simulation}
\label{sssec:movement_simulate}

Human activities in an apartment happen in shared public space (\emph{e.g.}, living room) and also in private space (\emph{e.g.}, master and second room).
In a functional residential building, the living room should well-connect the rooms and the main entrance~\cite{michalek_2002_architectural}, while furniture arrangement is a key factor that affects the resident movements in other rooms~\cite{yu_2011_make}.
Hence, this work decomposes the human-activity map into (i) room-room and room-entrance connections in the living room; and (ii) resident movements in the other rooms.

Figures~\ref{fig:activity_map}(c, d) depict the procedure of activity decomposition and movement simulation.
First, we identify locations of the main entrance of the apartment and entrances to all the other rooms from the floorplans in the RPLAN dataset~\cite{wu_2019_data-driven} and automatically connect them all together into a bi-directional graph (Figure~\ref{fig:activity_map}(c-top)).
Based on the connectivity graph, we adopt bi-RRT~\cite{lavalle_1998_rapidly-exploring} to simulate the resident movements in the living room (Figure~\ref{fig:activity_map}(d-top)) and produce a single-channel image, in which the pixel values indicate the probabilistic human activity \major{intensity, not activity types,} in the living room.
Similarly, we identify the entrance-furniture connections (Figure~\ref{fig:activity_map}(c-bottom)) and simulate the resident movements in the other rooms (Figure~\ref{fig:activity_map}(d-bottom)). 
Next, we merge the simulation results in the living room and other rooms together, yielding the final human-activity map (Figure~\ref{fig:activity_map}(e)).
We empirically assign weights of 6:4 to the simulation image in the living room and that in the other rooms, since human activities are more frequent in the living room than in the other rooms.

\begin{figure}[t]
	\centering
	\includegraphics[width=0.495\textwidth]{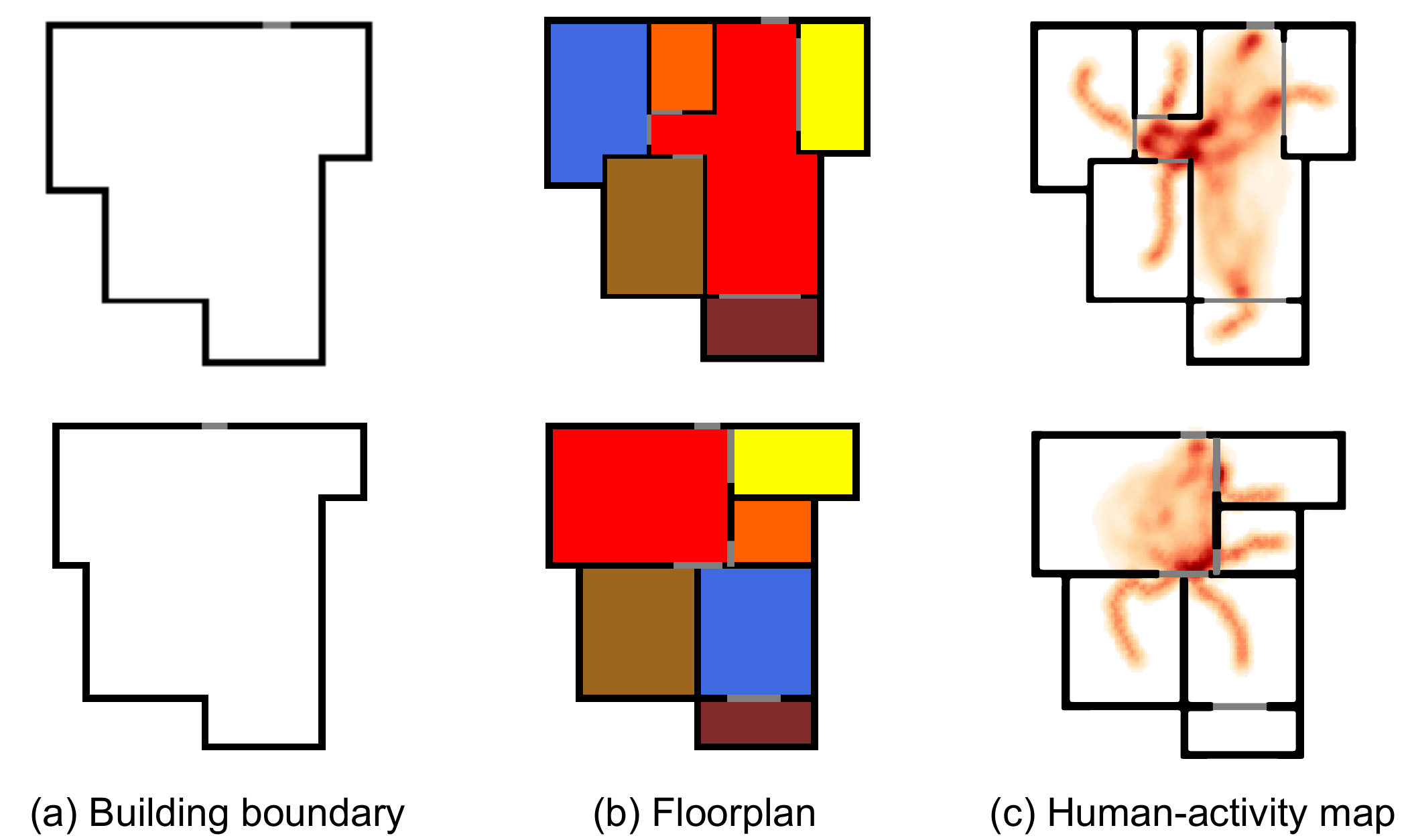} \\
	\vspace*{-2mm}
	\caption{Sample building boundaries (a) and floorplans (b) from the RPLAN dataset, and the corresponding human-activity maps by our simulation (c).}
	\label{fig:inputs}
	\vspace*{-2mm}
\end{figure}

In this way, we can produce a human-activity map for each floorplan in the RPLAN dataset.
Figure~\ref{fig:inputs} presents two example sets of building boundary, floorplan, and human-activity map, where the left two columns are samples from the RPLAN dataset, and the right-most one shows our simulated results.
Here, the two boundaries in Figure~\ref{fig:inputs}(a) are similar, whilst their corresponding floorplans in Figure~\ref{fig:inputs}(b) are quite different.
It is a challenging task for a network to recognize the small difference in the inputs and predicts diverse floorplans as the outputs.
We address the challenge by incorporating the human-activity map (see examples in Figure~\ref{fig:inputs}(c)) as guidance to help the network learn the mapping from building boundary to realistic floorplan.

\subsection{Generation from Building Boundary}
\label{ssec:activity_map_from_building_boundary}

So far, human-activity maps are synthesized from fine-grained real floorplans, which are available only in the training stage.
Since the test stage also needs human-activity maps to guide the floorplan generation, we thus need to generate human-activity maps from unseen boundaries.
We develop two alternative approaches using either an end-to-end GAN model (Sec.~\ref{sssec:automatic}) or a user interface (Sec.~\ref{sssec:semi-automatic}).

\begin{figure}[!t]
	\centering
	\includegraphics[width=0.495\textwidth]{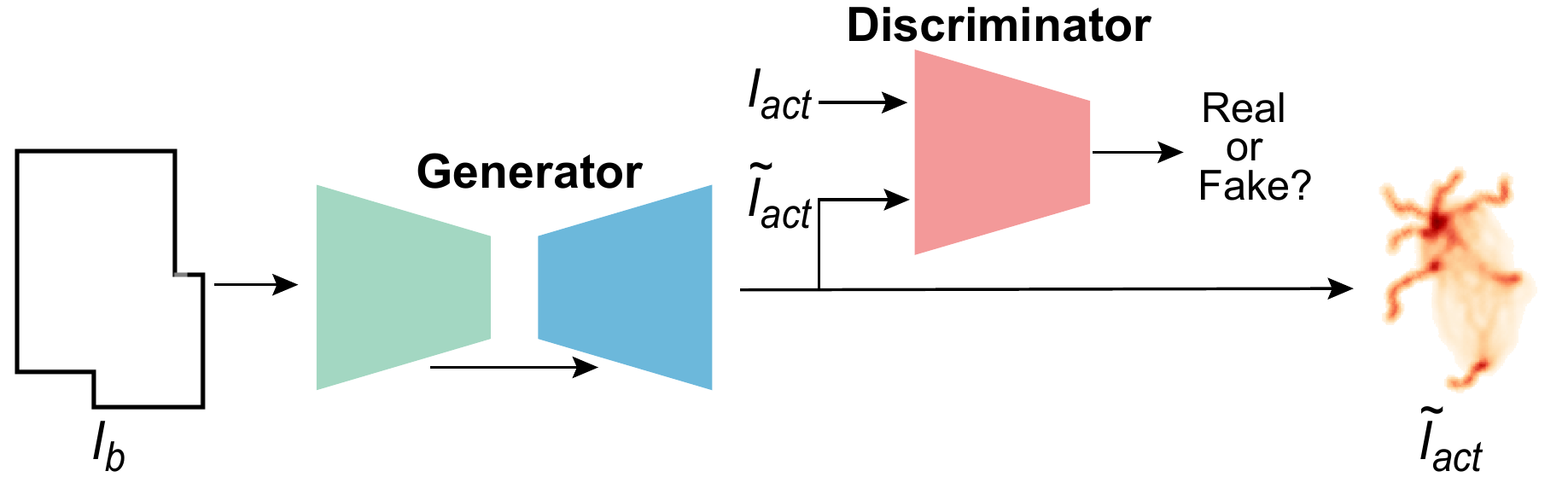} \\
	\vspace{-2mm}
	\caption{Architecture of the generative network for automatic human-activity map generation from building boundary.}
	\vspace{-2mm}
	\label{fig:activity_mask_net}
\end{figure}

\subsubsection{Automatic Approach with Generative Network}
\label{sssec:automatic}

We train a generative network to automatically generate human-activity maps from input building boundaries.
The training set consists of building boundary $I_b$ retrieved from the RPLAN dataset, and human-activity map $I_{act}$ synthesized from the associated floorplan paired with the building boundary.
Figure~\ref{fig:activity_mask_net} presents the overall network model, which is based on DCGAN~\cite{radford_2015_unsupervised}.

The generator learns to produce samples of human-activity map $\tilde{I}_{act}$ as the training data.
The standard DCGAN generates an image from a sampled latent vector in a uniform distribution.
However, this work requires inferring the latent vector from the input building boundary, rather than from a random sample.
Hence, we design the generator as an encoder-decoder network, of which the encoder converts the input building boundary $I_b$ into a latent vector, and the decoder maps the latent vector into a sample of human-activity map $\tilde{I}_{act}$.
Further, we add skip connections between each layer $i$ in the encoder and layer $n-i$ in the decoder to share low-level information between the encoder and the decoder.
Altogether, there are five layers for both the encoder and the decoder.
\tvcg{
We incorporate noise in the form of dropouts in the first three layers of the decoder at both training and test time, following the same practice as~\cite{isola_2016_Image-To-Image}, to enable stochastic outputs.
}

The discriminator learns to determine whether a given sample (human-activity map) is produced by a GAN model or synthesized from a real floorplan inside the RPLAN dataset using bi-RRT.
The discriminator has five layers to progressively down-sample the input image, similar to the encoder of the generator module.
We leverage an adversarial loss that can jointly optimize both the generator and the discriminator simultaneously; the adversarial loss is detailed in Sec.~\ref{ssec:from_boundary}.
Previous approaches have found it beneficial to mix the GAN objective with a more traditional loss, so we also use an L1 loss as in~\cite{isola_2016_Image-To-Image} for robustness.

\begin{figure}[t]
	\centering
	\includegraphics[width=0.495\textwidth]{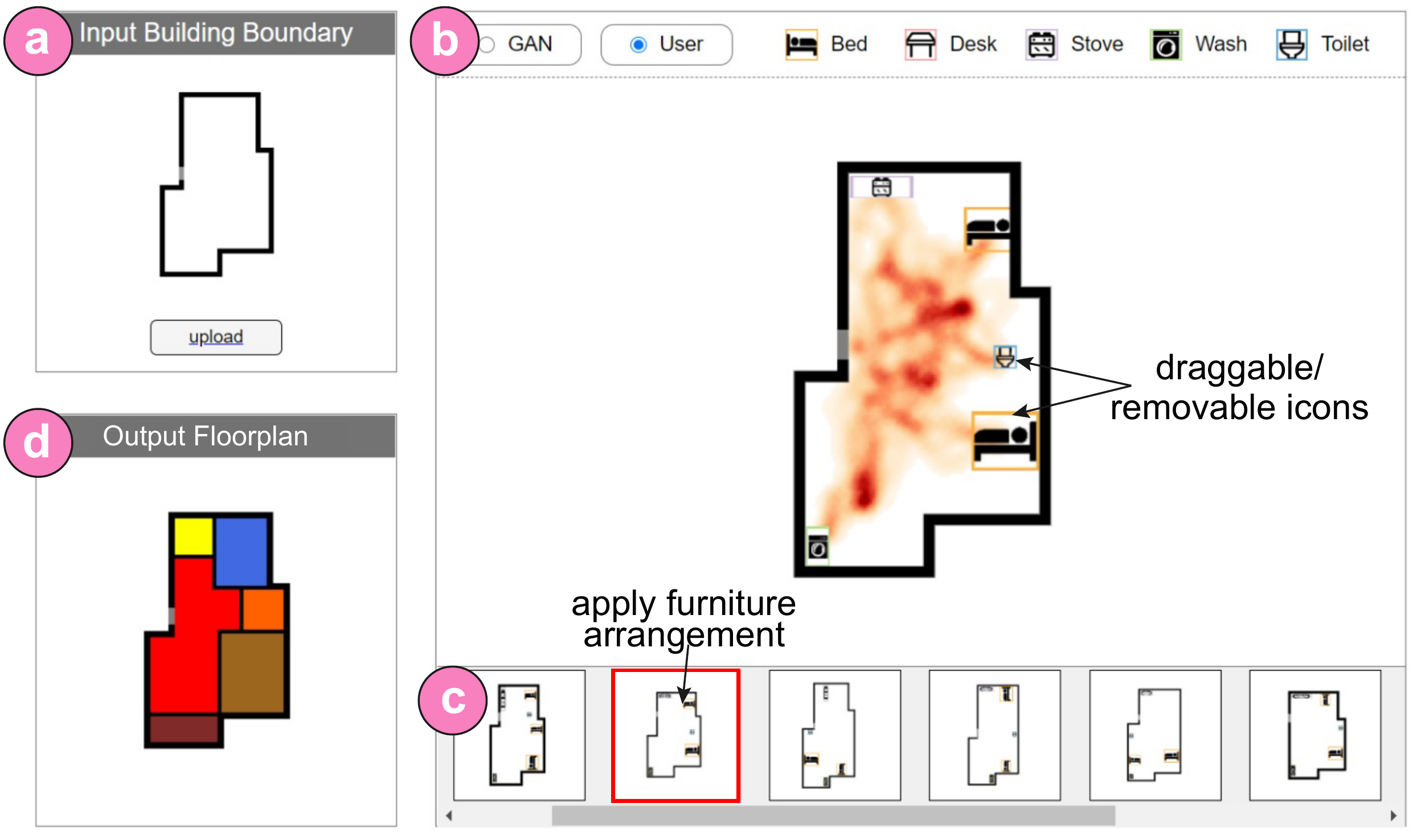}
	\vspace{-4mm}
	\caption{Our interface for semi-automatic human-activity map generation includes the \emph{Activity Control Panel} (b) and a \emph{Recommendation Gallery} (c). 
	From the input building boundary
	(a), our ActFloor-GAN generates a floorplan (d) under the guidance of the human-activity map shown in (b).
	}
	\vspace{-4mm}
	\label{fig:interface}
\end{figure}

\begin{figure*}[ht]
	\centering
	\includegraphics[width=0.995\textwidth]{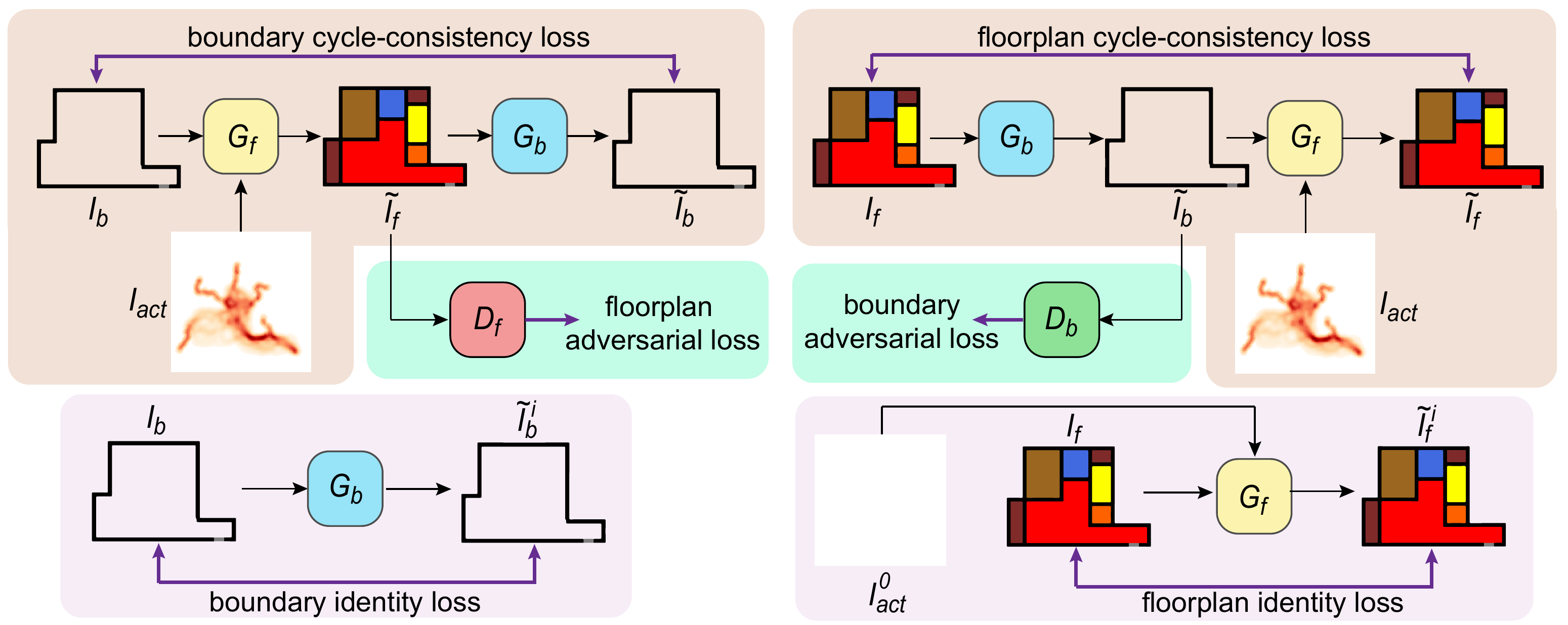} \\
	\vspace*{-2mm}
	\caption{
	Our ActFloor-GAN framework consists of two parts: (left) the one to learn from building boundaries, and (right) the other to learn from floorplans.
	Each part involves three losses: adversarial loss, cycle-consistency loss, and identity loss. $G_{f}$ and $G_{b}$ denote the generators that produce floorplan and building boundary, respectively, while $D_{f}$ and $D_{b}$ are the discriminators that determine whether the generated images are real floorplan or real building boundary, respectively. 
	}
	\label{fig:network_arch}
	\vspace*{-4mm}
\end{figure*}

\subsubsection{Semi-Automatic Approach with Interactive Interface}
\label{sssec:semi-automatic}
Though being fully automatic, the generative network offers no control over the floorplan generation, failing to offer design options that are preferred by architects.
To this end, we develop an interactive interface that allows users to manipulate furniture placement and to generate multiple human-activity maps, thus enabling the generation of diverse floorplans.
As shown in Figure~\ref{fig:interface}, the interface mainly consists of an \emph{Activity Control Panel} (Figure~\ref{fig:interface}(b)) and a \emph{Recommendation Gallery} (Figure~\ref{fig:interface}(c)).

The \emph{Activity Control Panel} provides two alternative approaches to generate human-activity map:
(i) users can choose to use the automatic approach described in Sec.~\ref{sssec:automatic}, by selecting the \emph{GAN} radio button on the top; and (ii) users can choose to manually manipulate the furniture placement and generate human-activity maps by selecting the \emph{User} radio button.
In the manual mode, users can \emph{add} furniture by dragging furniture icons into the building boundary, \emph{remove} furniture by right-clicking on the icons, or \emph{adjust} the furniture positions by dragging the furniture icons.
We synthesize a room entrance for each piece of furniture added into the apartment according to the furniture placement rules described in Sec.~\ref{sssec:furniture_place} and the empirical door placement rules described in~\cite{wu_2019_data-driven}. 
Note, room entrances are omitted in the figure, for simplicity.
Finally, we synthesize a human-activity map based on the positions of the main entrance, furniture, and room entrances using bi-RRT; see Sec.~\ref{sssec:movement_simulate} for details.

To alleviate the manual add and drag operations, we retrieve building boundaries from the RPLAN dataset that are similar to the input one, and display the top ten most similar ones in the \emph{Recommendation Gallery}.
Here, we compute the shape similarity between building boundaries using the Hu moment invariant~\cite{1057692}.
Users can select one of the building boundaries in the recommendation list, then apply furniture placement to the input building boundary. \major{Users are allowed to further add/remove furniture, and adjust furniture placement by dragging relevant icons.}
Besides, the interface also displays the input building boundary (Figure~\ref{fig:interface}(a)), and outputs the floorplan (Figure~\ref{fig:interface}) generated by our ActFloor-GAN under the guidance of the human-activity map presented in Figure~\ref{fig:interface}(b).

With the interface, users can create multiple feasible human-activity maps from the same building boundary.
Figure~\ref{fig:diversity} presents examples of diverse floorplans generated from different human-activity maps on the same boundary.
\section{ActFloor-GAN}
\label{sec:floorplan_generation}

We design ActFloor-GAN to predict floorplans from building boundaries.
Our goal is to learn a mapping function from the domain of building boundaries $P_{data}(I_b)$ to the domain of floorplans $P_{data}(I_f)$, through the guidance of samples from the human-activity map domain $P_{data}(I_{act})$.
ActFloor-GAN is a new deep framework that takes an input image $I_{b} \in P_{data}(I_b)$ that represents the building boundary and a human-activity map $I_{act} \in P_{data}(I_{act})$ to guide the floorplan generation via our re-formulated cycle-consistency constraints.
Figure~\ref{fig:network_arch} outlines the overall network architecture of ActFloor-GAN, which has two parts: one to learn from the building boundaries to floorplans $G_f: P_{data}(I_b) \rightarrow P_{data}(I_f)$ (Sec.~\ref{ssec:from_boundary}), and the other to learn from the floorplans to building boundaries $G_b: P_{data}(I_f) \rightarrow P_{data}(I_b)$ (Sec.~\ref{ssec:from_floorplan}).

\subsection{Learning from Building Boundary}
\label{ssec:from_boundary}
Figure~\ref{fig:network_arch}(left) shows the part of our framework that learns to generate floorplans from building boundaries.
Given an image of boundary $I_{b}$, we first use generator $G_{f}$ to produce a floorplan $\tilde{I}_f$.
As discussed above, we aim to produce multiple floorplans from the same building boundary, by taking human-activity map $I_{act}$ as guidance to constrain the design generation.
Hence, generator $G_f$ also takes $I_{act}$ as a part of its inputs, \emph{i.e.}, $\tilde{I}_f = G_f(I_b, I_{act})$.
Then, a discriminator $D_{f}$ is employed to distinguish if $\tilde{I}_f$ is a real floorplan or not, \emph{i.e.}, $D_f(\tilde{I}_f) =$ $real$ or $fake$?
We use a \emph{floorplan adversarial loss} to optimize both generator $G_{f}$ and discriminator $D_f$~\cite{mirza_2014_Conditional}.

\vspace{-2mm}
\begin{equation}
\label{eqn:b001}
\begin{split}
L^b_{GAN}(G_f, \, &D_f) = \mathbb  {E}_{I_f \sim P_{data}(I_f)}[\log D_f(I_f)]\\
& + \mathbb{E}_{I_b \sim P_{data}(I_b)}[\log(1-D_f(G_{f}(I_b, I_{act})))]
\end{split}
\end{equation}

\noindent
where $\mathbb{E}$ indicates the error; $P_{data}(I_f)$ and $P_{data}(I_b)$ indicate the data distribution of $I_f$ and $I_b$, respectively; and $\ast \sim P_{data}(\ast)$ indicates that $\ast$ is selected from the respective dataset $P_{data}(\ast)$.

To prevent generator $G_{f}$ from cheating discriminator $D_{f}$ too easily, we design another generator $G_{b}$ that produces a boundary image $\tilde{I}_b$ from the generated floorplan image $\tilde{I}_f$, \emph{i.e.}, $\tilde{I}_b = G_b(\tilde{I}_f)$.
We leverage a \emph{boundary cycle-consistency loss} to optimize the network~\cite{zhu_2017_cyclegan}:

\vspace{-2mm}
\begin{equation}
\label{eqn:b002}
\begin{split}
L^b_{cycle}(G_{f}, G_b) = 
\mathbb{E}_{I_b \sim P_{data}(I_b)}[\Vert G_b(G_{f}(I_{b}, I_{act})) - I_{b} \Vert _1]
\qquad\qquad
\end{split}
\end{equation}

\noindent
where $\Vert \cdot \Vert _1$ denotes the L1 loss of the difference between the two images on corresponding pixel.

Last, we use the real boundary $I_b$ as the input of $G_{b}$, and generate a building boundary $\tilde{I}^i_b$ similar to $I_b$, \emph{i.e.}, $\tilde{I}^i_b = G_b(I_b)$.
We would like to force the generated building boundary $\tilde{I}^i_b$ to be the same with $I_b$, to make sure the generated floorplan is within the same boundary as the input building boundary.
We leverage a \emph{boundary identity loss} to achieve the goal:

\vspace{-2mm}
\begin{equation}
\label{eqn:b006}
\begin{split}
L^b_{identity}(G_{b}) =
\mathbb {E}_{I_b \sim P_{data}(I_b)} [\Vert G_{b}(I_b) - I_b \Vert _1].
\end{split}
\end{equation}

\subsection{Learning from Floorplan}
\label{ssec:from_floorplan}

Figure~\ref{fig:network_arch}(right) shows the other part of our framework that learns to generate building boundaries from floorplans.
Given a floorplan image $I_{f}$, we first use generator $G_{b}$ to produce a boundary image $\tilde{I}_b$, \emph{i.e.}, $\tilde{I}_b = G_b(I_f)$.
Then, we employ discriminator $D_{b}$ to distinguish if $\tilde{I}_b$ is a real boundary or not, \emph{i.e.}, $D_b(\tilde{I}_b) =$ $real$ or $fake$?
We leverage the \emph{boundary adversarial loss} to supervise the boundary generation in the generative adversarial game:

\vspace{-2mm}
\begin{equation}
\label{eqn:f007}
\begin{split}
L^f_{GAN}(G_b, D_b)& =\mathbb {E}_{I_b \sim P_{data}(I_b)}[\log D_b(I_b)]\\
& + \mathbb{E}_{I_f \sim P_{data}(I_f)}[\log(1-D_b(G_{b}(I_f)))].
\end{split}
\end{equation}

Similarly, in order to prevent generator $G_{b}$ from cheating discriminator $D_{b}$ too easily, we design another generator $G_{f}$ whose inputs are the generated boundary image $\tilde {I}_b$ and a human-activity map $I_{act}$, and output is a floorplan image $\tilde {I}_f$, \emph{i.e.}, $\tilde {I}_f = G_f(\tilde {I}_b, I_{act})$. 
We leverage \emph{floorplan cycle-consistency loss} to train the network~\cite{zhu_2017_cyclegan}:

\vspace{-2mm}
\begin{equation}
\label{eqn:f010}
\begin{split}
L^f_{cycle}(G_{b}, G_{f}) = \mathbb {E}_{I_f \sim P_{data}(I_f)}[\Vert G_{f}(G_b(I_f), I_{act}) - I_f\Vert _1].
\end{split}
\end{equation}

Last, we use an empty human-activity map $I^0_{act}$ with all zero values and the real floorplan $I_f$ as the input to $G_{f}$.
Ideally, the generated floorplan $\tilde{I}^i_f$ should be identical to the real floorplan $I_f$, by minimizing the \emph{floorplan identity loss}:

\vspace{-2mm}
\begin{equation}
\label{eqn:f013}
\begin{split}
L^f_{identity}(G_{f}) =
\mathbb {E}_{I_f \sim P_{data}(I_f)}[\Vert G_{f}(I_f, I^0_{act}) - I_f \Vert _1].
\end{split}
\end{equation}

\major{Note that the generator $G_{b}$ could be trivially replaced by a non-learning method that extracts the boundary of a given floorplan.
We keep $G_{b}$ in the training process, so that the generator $G_{f}$ can be trained synchronously to better learn the mapping from boundary to floorplan.
See Sec.~\ref{ssec:quantitative} for a comparison with prior methods, especially \emph{Pix2Pix} without using the cycle-consistency constraint, showing that the constraint can help enhance the results.
}

\subsection{Loss Function and Training Strategy}
\label{ssec:loss}
Putting all the losses in the two parts together, the overall loss function for training our ActFloor-GAN model is
\begin{equation}
\label{eqn:f014}
\begin{split}
& L_{final}(G_{b},G_{f},D_{b},D_{f}) =
\\
& \lambda_1 (L^b_{GAN}(G_f,D_f) + L^f_{GAN}(G_b,D_b))
\\
& + \lambda_2 (L^b_{cycle}(G_f, G_b) + L^f_{cycle}(G_b, G_f))
\\
& + \lambda_3 (L^b_{identity}(G_b) + L^f_{identity}(G_f)).\\
\end{split}
\end{equation}

Following the empirical settings as in~\cite{mirza_2014_Conditional}, we achieved good experimental results by setting weights $\lambda_1$ , $\lambda_2$, and $\lambda_3$ as 1, 10, and 5, respectively.

We adopt the network architecture designed by Johnson et al.~\cite{Johnson_2016_Perceptual} as our generator network.
Input to generator ${G_b}$ is a floorplan image in three-channel RGB format, while that for generator ${G_f}$ is a four-channel grayscale image by concatenating a three-channel building boundary image with a one-channel human-activity map.
We use PatchGAN~\cite{isola_2016_Image-To-Image} to help discriminators ${D_b}$ and ${D_f}$ distinguish whether the image patches are real or fake.
Building boundaries and floorplans are derived from the RPLAN dataset~\cite{wu_2019_data-driven}, while human-activity maps are synthesized from the floorplans as described in Sec.~\ref{ssec:activity_map}.
There are in total 80K pairs of building boundaries, floorplans, and human-activity maps, and we randomly split them into 75k-2.5k-2.5k for training-validation-test sets. 
We set the batch size to 1 and employed Adam optimizer~\cite{Kingma_2015_AdamAM} ($\beta_{1}$ = 0.5, $\beta_{2}$ = 0.999) with a fixed learning rate of 0.0002.
The network architecture was implemented in PyTorch and ran on a server equipped with an NVIDIA GTX 1080Ti GPU card.

\begin{figure}[t]
	\centering
	\includegraphics[width=0.495\textwidth]{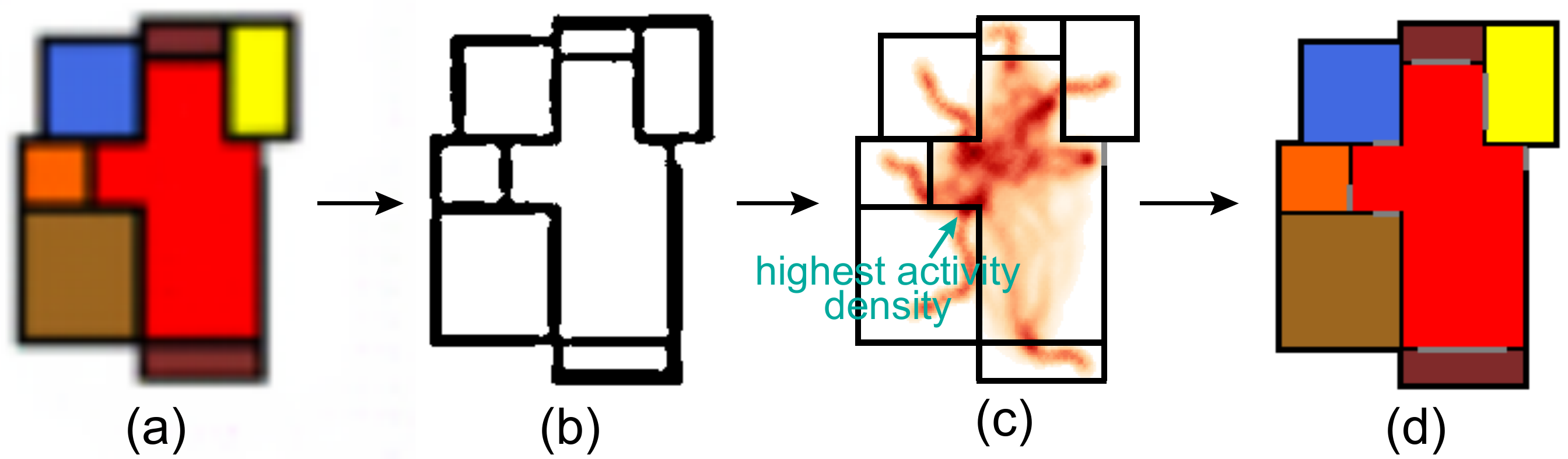} \\
	\vspace*{-2mm}
	\caption{Illustration of the floorplan vectorization process. (a) Raster image with pixel colors indicating rooms and walls. (b) Noisy exterior and interior walls. (c) Vectorized walls and the corresponding human-activity map. (d) Final vector floorplan with room and door labels.}
	\label{fig:post-processing}
	\vspace*{-4mm}
\end{figure}

\subsection{Floorplan Vectorization}
\label{ssec:vector}

The output of ActFloor-GAN is a three-channel image with pixel colors indicating the room types and positions, as well as the wall positions (Figure~\ref{fig:post-processing}(a)).
Further, we leverage a post-processing module to convert the raster image into a vectorized floorplan, to make the results usable by architects.
Figure~\ref{fig:post-processing} illustrates the process.
From a raster image (Figure~\ref{fig:post-processing}(a)), we first binarize it based on adaptive thresholding to obtain the exterior and interior walls (Figure~\ref{fig:post-processing}(b)).
The walls are, however, rather noisy and discontinuous.
We further perform morphological closing operations on the results, yielding vertical and horizontal straight lines that form closing boxes.
We assign semantics of room types to each room according to the predictions.
Next, we \tvcg{find the position with the highest activity density in each room (except for the living room)} to position the internal doors.
Finally, the vectorized floorplan (Figure~\ref{fig:post-processing}(d)) is generated.
\section{Evaluation and Discussion}
\label{sec:evaluation}

To evaluate the ActFloor-GAN model, we \TVCG{examined the performance of the GAN model for automatic human-activity map generation (Sec.~\ref{ssec:assess_activity_gan})}, conducted a quantitative comparison between ActFloor-GAN and \TVCG{four} \major{prior methods} (Sec.~\ref{ssec:quantitative}), and also performed a user evaluation by architects to compare our generated floorplans with those generated by \emph{Wu et al.}~\cite{wu_2019_data-driven} and those from the RPLAN dataset (Sec.~\ref{ssec:qualitative}).
Then we demonstrate the capability of our approach in generating diverse floorplans (Sec.~\ref{ssec:diversity}), followed by a discussion in the end (Sec.~\ref{ssec:discuss}).

\TVCG{
\subsection{Assessment of Automatic Activity Map Generation}
\label{ssec:assess_activity_gan}

We use mutual information (MI)~\cite{maes_1997_multimodality}, a concept from the information theory, to assess the GAN model for automatic human-activity map generation.
The metric has been applied in medical imaging (\eg\cite{jenkinson2002improved}) and data visualization (\eg\cite{zeng_2019_raeb}).
Notice that MI computation requires a preliminary step of image registration, which can be skipped in this work since the input and output images of the GAN model are naturally aligned.
Also, notice that the human-activity maps can be seen as discrete random variables that take intensity values in the range [0, 255].
As such, we can measure MI between a ground-truth human-activity map $I_{act}$ and a synthesized one $\tilde{I}_{act}$ as

\vspace{-2mm}
\begin{equation}
	\label{eqn:nmi001}
	\begin{split}
		MI(I_{act}, \tilde{I}_{act})=\sum_{x \in h(I_{act})} \sum_{y \in h(\tilde{I}_{act})} p(x, y) log(\frac{p(x, y)}{p(x)p(y)}), 
	\end{split}
\end{equation}
where $h(\cdot)$ denotes the normalized intensity histogram of an image, $p(\cdot)$ is the marginal probability distribution function, and $p(x, y)$ is the joint probability function of the intensity histograms.
To eliminate the impacts of image size, we use a normalized version of MI (denoted as NMI):

\vspace{-2mm}
\begin{equation}
	\label{eqn:nmi002}
	\begin{split}
		NMI(I_{act}, \tilde{I}_{act})=\frac{2MI(I_{act}, \tilde{I}_{act})}{H(I_{act}) + H(\tilde{I}_{act})},
	\end{split}
\end{equation}
where $H(I_{act}) = - \sum_{x \in h(I_{act})} p(x) log(p(x))$ measures the entropy of the normalized intensity histogram.
NMI ranges in [0, 1], where 0 indicates no mutual information while 1 indicates perfect correlation.

\begin{figure}[t]
	\centering
	\includegraphics[width=0.495\textwidth]{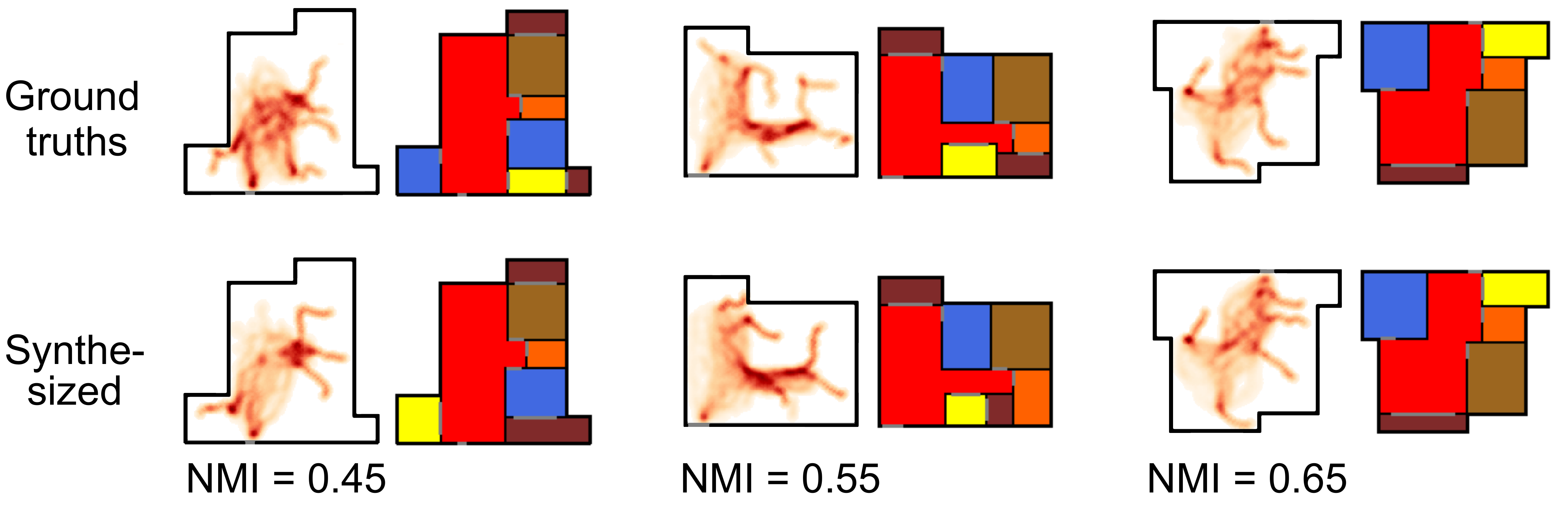} \\
	\vspace*{-2mm}
	\caption{
		\TVCG{
		Visual difference between ground truths (top) and our synthesized human-activity maps (bottom) of the GAN model for automatic human-activity map generation.
		The corresponding ground-truth and synthetic floorplans are presented besides, while the NMI values are at the bottom, showing that quality of the synthetic floorplans are highly dependent on the similarity of the human-activity map.
		}
	}
	\label{fig:compare_map}
	\vspace*{-4mm}
\end{figure}

We train the GAN model with 75k pairs of building boundaries and simulated human-activity maps, and evaluate the performance using 2.5k testing pairs.
The average NMI value is 0.565, and the minimum and maximum values are 0.434 and 0.669, respectively.
Figure~\ref{fig:compare_map} presents some examples of ground-truth and synthesized human-activity maps, sorted by their NMI values in ascending order.
Notice that the ground-truth and synthesized activity maps look more similar for larger NMI values.
The synthetic floorplans guided by the human-activity maps are presented on the right sides correspondingly.
It is also observed that the floorplans become more similar when the NMI value is large.
For the floorplans shown on the left (NMI = 0.45), the number of rooms and room types of the synthesized floorplans are different than those of the ground truths.
In contrast, for the floorplans on the right (NMI = 0.65), the topological and geographical properties of the synthesized floorplans are identical to those of the ground truths.
}

\subsection{\major{Comparison with Prior Methods}}
\label{ssec:quantitative}

The quality of the generated floorplans depends much on the network predictions.
It is necessary to evaluate the prediction results with \major{prior methods} that can predict pixel-wise room types from input boundaries.
Here, we omit \emph{Wu et al.}~\cite{wu_2019_data-driven}, which predicts room centers instead of piecewise room types, and other approaches (\eg\cite{nauata_2020_house-gan}) that require additional inputs such as room adjacency relations.

\vspace{1mm}
\noindent
\emph{\major{Prior methods}}.
Some attempts, \eg ArchiGAN~\cite{Chaillou_2019_archigan}, directly predict pixel-wise room types using \emph{Pix2Pix}~\cite{isola_2016_Image-To-Image}.
Also, \emph{CycleGAN}~\cite{zhu_2017_cyclegan}, which was originally designed for general image-to-image translation, can also be employed to predict pixel-wise room types.
\tvcg{We also compare with a network architecture that is the same as ActFloor-GAN but without guidance by the human-activity maps, denoted as \emph{Floor-GAN}.}
We further compare with \emph{Graph2Plan}~\cite{hu_2020_Graph2Plan} that predicts room boxes and a raster image of room types, \major{using the ground-truth room graph as guidance}.
We adopted author-provided implementations and settings and re-trained the models on the RPLAN dataset.
Since the \major{prior methods} are fully automatic, we utilize automatically-generated human-activity maps to guide ActFloor-GAN (denoted as \emph{Our}).

\begin{figure}[t]
	\centering
	\includegraphics[width=0.495\textwidth]{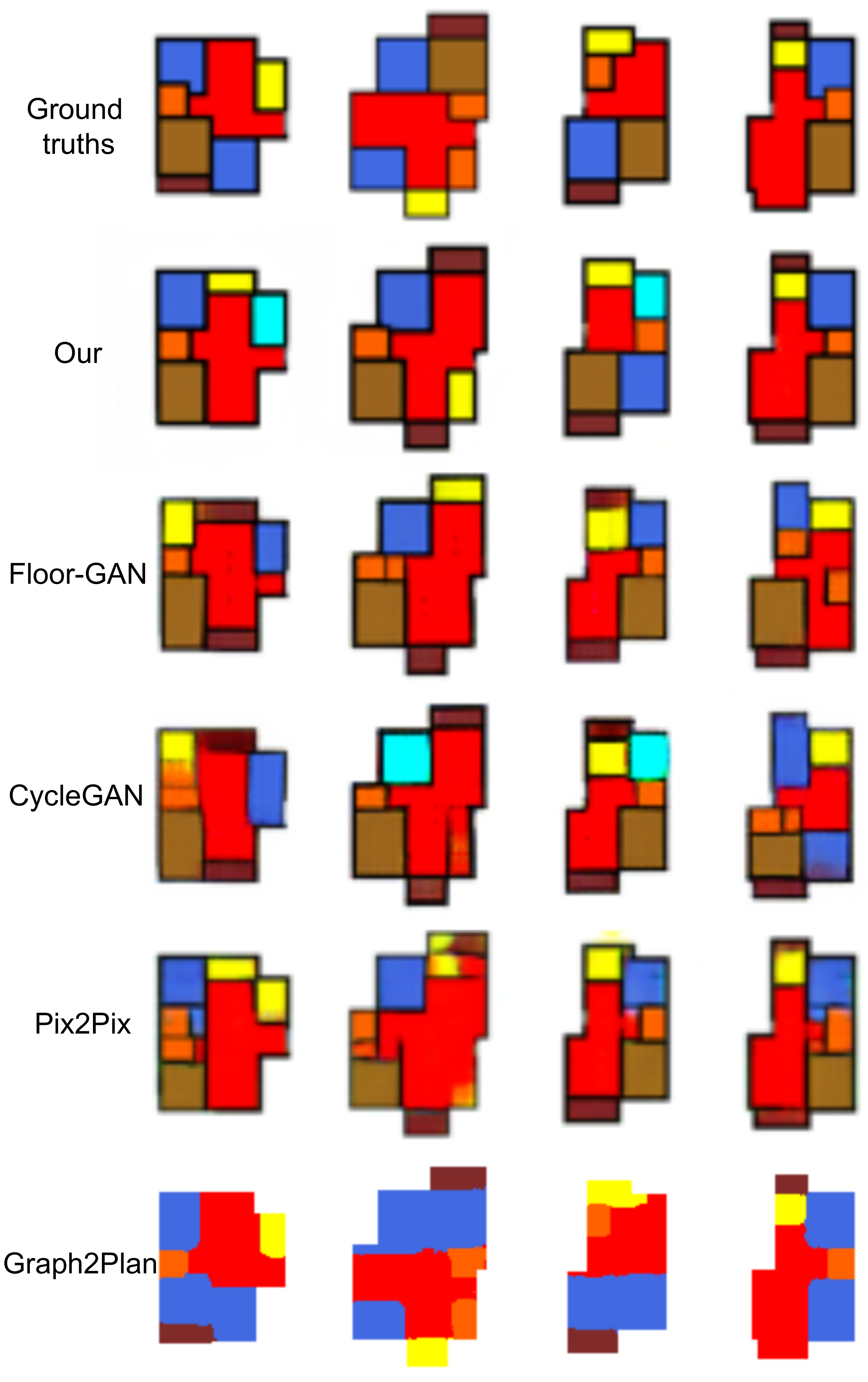} \\
	\vspace*{-2mm}
	\caption{Visual comparisons with competitors. From top to bottom: ground truths, predictions by Our, Floor-GAN, CycleGAN~\cite{zhu_2017_cyclegan}, Pix2Pix~\cite{isola_2016_Image-To-Image}, and \TVCG{Graph2Plan~\cite{hu_2020_Graph2Plan}}.}
	\label{fig:compare_models}
	\vspace*{-4mm}
\end{figure}

\begin{figure}[t]
	\centering
	\includegraphics[width=0.495\textwidth]{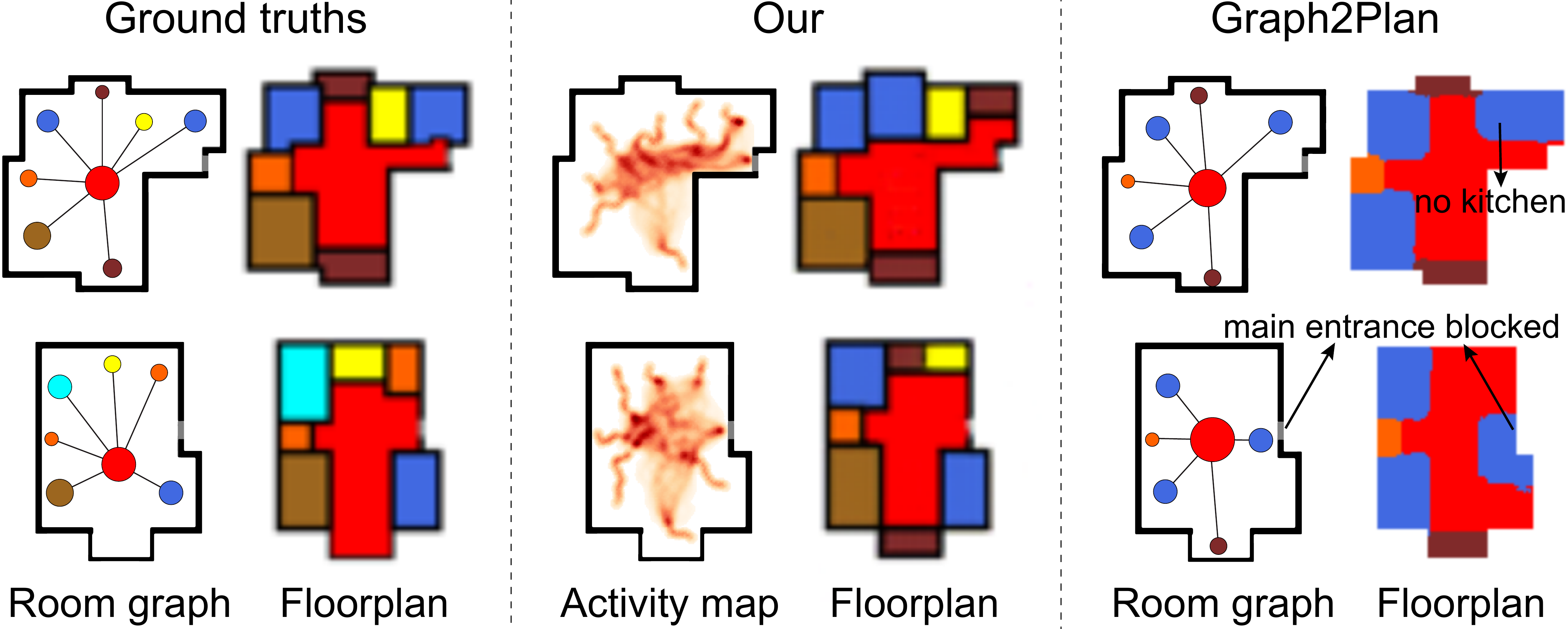} \\
	\vspace*{-2mm}
	\caption{\major{Comparisons of \emph{Our} guided by automatically synthesized human-activity maps, and \emph{Graph2Plan} guided by room graphs of the most similar floorplans as the input boundaries retrieved from the dataset.}}
	\label{fig:fig_vs_g2p}
	\vspace*{-4mm}
\end{figure}

\vspace{1mm}
\noindent
\emph{Qualitative comparison}.
Figure~\ref{fig:compare_models} presents visual comparisons with the competitors.
The top row presents ground truths (denoted as \emph{GT}) from the RPLAN dataset, and the other rows present results by \emph{Our}, \tvcg{\emph{Floor-GAN}}, \emph{CycleGAN}, \emph{Pix2Pix}, \TVCG{and \emph{Graph2Plan}}, respectively.
\TVCG{For fairness, all results presented here are intermediate raster images without vectorization.}

Overall, \emph{Our} model produces the most similar results with \emph{GT}, whilst \emph{Pix2Pix} produces noisy results, and \emph{CycleGAN} and \tvcg{\emph{Floor-GAN}} produce grid rooms but the resulting room types differ from \emph{GT}.
Here, the \emph{Pix2Pix} model employs the fewest losses in comparison with the other models, yielding non-closing rooms and unqualified predictions for the vectorization module.
\emph{CycleGAN} recruits an additional cycle-consistency loss and \tvcg{\emph{Floor-GAN} further adds the identity loss}, to constrain the network, thus their predictions are finer with the grid rooms.
However, it is difficult for the models to distinguish similar building boundaries.
For example, the input building boundaries in columns 3 and 4 are very similar, and the \emph{Floor-GAN} result in column 3 is very similar to the \emph{GT} in column 4.
\tvcg{This indicates that cycle-consistency and identity losses alone are susceptible to adversarial perturbations of input boundaries.}
In contrast, \emph{Our} model can recognize the difference, benefiting from the guidance of automatically generated human-activity maps.
\emph{Graph2Plan} also produces fine results, but the method simplifies the prediction by categorizing the \emph{master room} (dark golden in GT), \emph{second room} (blue in GT), and \emph{study room} (cyan in GT) as \emph{bedroom} (blue in Graph2Plan).
Additional information is needed to distinguish these room types.
\major{Notice here that \emph{Graph2Plan} is guided by the ground-truth room graphs.
We observe some improper floorplans may be generated, \eg missing kitchen (Figure~\ref{fig:fig_vs_g2p} (top-right)) or main entrance blocked by non-living rooms (Figure~\ref{fig:fig_vs_g2p} (bottom-right)), using room graphs of the most similar floorplans as the input boundaries retrieved from the dataset.}

\begin{table}[h]
\centering
\scriptsize
\caption{Quantitative comparison with Pix2Pix, CycleGAN, Floor-GAN, and \TVCG{Graph2Plan} in terms of pixel-wise accuracy (MSE and MAE), and vectorization success rate.}
\begin{tabular}{c | c | c | c| c |c}
	\hline
	& Pix2Pix & \tvcg{CycleGAN} & Floor-GAN & \TVCG{Graph2Plan} & Our \\ \hline
MSE & 0.569 & \tvcg{0.183} & 0.176	& \TVCG{0.238} & \textbf{0.111} \\ \hline
MAE & 0.349 & \tvcg{0.166} & 0.160	& \TVCG{\textbf{0.079}} & 0.096 \\ \hline
\tvcg{Vector.} & \tvcg{17/100} & \tvcg{25/100} & \tvcg{27/100} & \TVCG{\textbf{100/100}} & \tvcg{83/100}\\ \hline
\end{tabular}
\vspace{-4mm}
\label{table:compare_models}
\end{table}

\vspace{1mm}
\noindent
\emph{Quantitative comparison}. \tvcg{We further conduct quantitative comparisons with the \major{prior} models, in terms of both the prediction accuracy and vectorization success rate.
Table~\ref{table:compare_models} presents the quantitative comparison results.}

\begin{itemize}

\item
We compute MSE and MAE between the ground truths and the results by \emph{Pix2Pix}, \emph{CycleGAN}, \tvcg{\emph{Floor-GAN}}, \TVCG{\emph{Graph2Plan}}, and \emph{Our}.
Smaller MSE and MAE values indicate better performances.
Table~\ref{table:compare_models} shows that \emph{Our} model achieves \TVCG{better} performance \TVCG{than \emph{Pix2Pix}, \emph{CycleGAN}, and \emph{Floor-GAN} that predict the same number of room types as \emph{Our}}, as both our MSE and MAE values are the smallest.
\TVCG{\emph{Graph2Plan} achieves the smallest MAE but a higher MSE than \emph{Our}}.
We admit that the floorplan generation task may not have a unique floorplan solution for the same boundary, so the use of the ground truths in this experiment is mainly to show the effectiveness of our framework over the \major{prior methods} in terms of capturing the latent geometry and topology properties of room layouts.

\item
We further check whether pixel-wise predictions by \emph{Our} and by the \major{prior} models can be successfully vectorized (see Sec.~\ref{ssec:vector}).
A vectorization is regarded as a success if the following conditions are met:
(i) every room must form a closed polygon and pixels within each room must have a dominant room type;
(ii) the number of room types is balanced, \eg at least one living room and one master room;
and (iii) the main entrance and internal doors (except bathroom and balcony) are directly connected to the living room.
We randomly select 100 boundaries and corresponding predictions from \emph{Our} and \major{prior} models, and measure their vectorization success rates as presented in Table~\ref{table:compare_models}.
\emph{Our} model can successfully vectorize 83 out of 100 predictions, which is significantly higher \TVCG{than \emph{Pix2Pix}, \emph{CycleGAN}, and \emph{Floor-GAN}}.
\emph{Floor-GAN} succeeds only 27 times, which however is still higher than the other two models.
\TVCG{\emph{Graph2Plan} succeeds 100 times using predicted room boxes, since it adopts a different vectorization strategy from \emph{Our} and others.
The method requires additional networks for predicting room boxes, whilst \emph{Our} model is lightweight.
}

\end{itemize}

\begin{table}[h]
	\caption{Our questions focus on geometry (Q1 - Q2), topology (Q3 - Q4), and overall design (Q5) of floorplans.}
	\begin{tabular}{c|l}
		\hline
		\rule{0pt}{2ex}  
		Q1	& Which one has better designed \emph{room sizes}? \\ \cline{2-2}
		\rule{0pt}{2ex}  
		Q2 	& Which one has better designed \emph{room shapes and aspect ratios}? \\ \hline
		\rule{0pt}{2ex}  
		Q3	& Which one has better designed \emph{room orientations}? \\ \cline{2-2}
		\rule{0pt}{2ex}  
		Q4	& Which one has better designed \emph{room connections}? \\ \hline 
		\rule{0pt}{2ex}  
		Q5 	& Overall, which one is better designed? \\ \hline                 
	\end{tabular}
	\label{table:floorplan_questions}
\end{table}

\subsection{Subjective Evaluation by Architects}
\label{ssec:qualitative}

Next, we explore whether our generated floorplans satisfy the requirements of architects.
In consultation with our collaborating architect, we formulate five questions on the quality of floorplan design, as listed in Table~\ref{table:floorplan_questions}.
Here, \emph{Q1 - Q2} concern geometry, \emph{Q3 - Q4} reflects topology, and \emph{Q5} is the overall design.
We compare the floorplans by our method with those from the RPLAN dataset that are real-world floorplans, and those by \emph{Wu et al.}~\cite{wu_2019_data-driven} that require only the building boundary as the same with our approach.

\begin{figure}[t]
	\centering
	\includegraphics[width=0.485\textwidth]{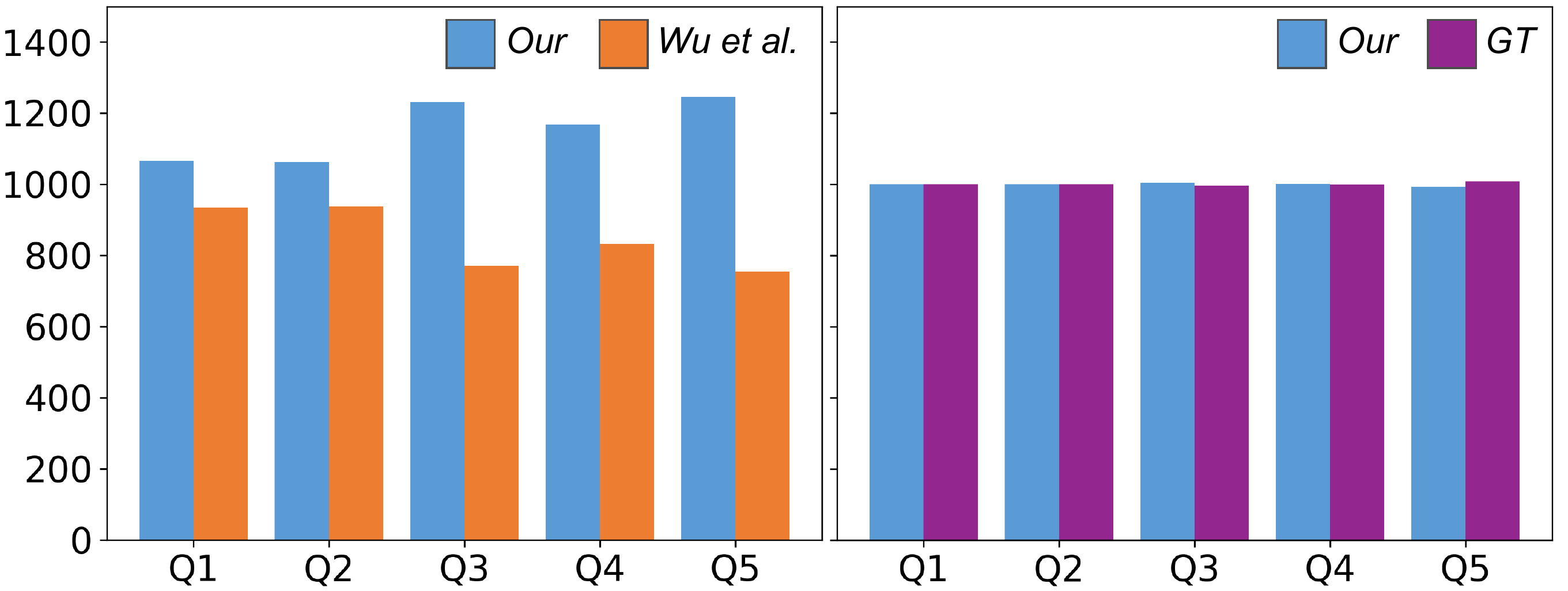} \\
	\caption{Elo ratings of \emph{Our} vs. \emph{Wu et al.} (left), and \emph{Our} vs. \emph{GT} (right) for \emph{Q1 - Q5}. \emph{Our} results are much better than \emph{Wu et al.} in terms of the topology-related criteria (Q3 \& Q4), and are of the compelling quality with the \emph{GT} that are professionally-designed floorplans.}
	\label{fig:userstudy}
	\vspace*{-3mm}
\end{figure}

\noindent
\emph{Participants}. We recruited five architects as participants in this study. 
All are graduate students holding a bachelor's degree in architecture.
They all have practical experience in designing floorplans for residential buildings.

\vspace{1mm}
\noindent
\emph{Procedure}.
We randomly selected 30 paired floorplans from the ground truths (\emph{GT}), the results by \emph{Our} model and also the results by \emph{Wu et al.}
We then form two sets of pairwise comparisons, \emph{Our} vs. \emph{GT}, and \emph{Our} vs. \emph{Wu et al.}
Examples of \emph{Our} vs. \emph{GT} are presented in Figure~\ref{fig:com_gt}.
We then asked the participants to make a comparison by giving a rating of ``better/worse/equally-good" for each pair of floorplans against \emph{Q1 - Q5}.
Notice that the participants were not aware of the origin of the floorplan: \emph{GT}, \emph{Our}, or \emph{Wu et al.}

\vspace{1mm}
\noindent
\emph{Result}.
We collected a total of 1500 answers (2 sets $\times$ 30 pairs $\times$ 5 participants $\times$ 5 questions).
For each set of comparison, we employed the Elo rating system to calculate the ratings for all five questions of the two players, \ie \emph{Our} vs. \emph{GT}, and \emph{Our} vs. \emph{Wu et al.}
The system predicts the outcome of a match between the two players by using an expected score formula, \eg a player whose rating is 100 points greater than their opponent's is expected to win 64\% of the time.
If Player $A$ has a rating of $R_A$ and Player $B$ has a rating of $R_B$, the expected scores for Players $A$ and $B$ are:
\begin{equation}
\label{eqn:e01}
\begin{split}
E_{A} = \frac{1}{1+10^{(R_B-R_A)/400}}, E_{B} = \frac{1}{1+10^{(R_A-R_B)/400}}.
\end{split}
\end{equation}

Suppose Player $A$ was expected to score $E_A$ points but actually scored $S_A$ points, the formula for updating its rating $\tilde R_A$ is:

\vspace{-4mm}
\begin{equation}
\label{eqn:e03}
\begin{split}
\tilde R_A = R_A+K(S_A-E_A)
\end{split}
\end{equation}

\vspace{-2mm}
\noindent
where $K$ is the maximum possible adjustment per game.
In this work, we set a constant value for $K$ as the number of players multiplied by 42, \ie 84.
For each question, a player is assigned an initial rating of 1000.
Every time a game is played, the change in the Elo rating of the players depends on the outcome and the expected outcome.
A win (``better'') counts a score of 1, loss (``worse'') counts a score of 0, and draw (``equally-good'') counts a score of 0.5.

Figure~\ref{fig:userstudy} presents the Elo ratings of \emph{Our} vs. \emph{Wu et al.} on the left, and \emph{Our} vs. \emph{GT} on the right, for Q1 - Q5.
From Figure~\ref{fig:userstudy}(left), our approach (\emph{Our}) produces much better floorplans than \emph{Wu et al.} in terms of topology-related criteria, \ie \emph{room orientation} (Q3) and \emph{room connection} (Q4), and slightly better results than \emph{Wu et al.} in terms of geometry-related criteria, \ie \emph{room size} (Q1) and \emph{room shape and aspect ratio} (Q2).
The reason is probably that \emph{Wu et al.} focuses on learning the geometry properties, specifically the center of room positions, in the RPLAN dataset, whilst neglecting the topological relations among rooms.
In contrast, the room topology information is embedded in human-activity maps, so ActFloor-GAN can effectively infer the topology information.
The difference promotes more preference for our results in terms of the \emph{overall design} (Q5). 
From Figure~\ref{fig:userstudy}(right), our results receive almost the same Elo ratings as those of \emph{GT} for all Q1 to Q5, indicating that ActFloor-GAN can produce floorplans of the compelling quality with the professionally-designed floorplans.

\subsection{Diverse Floorplans}
\label{ssec:diversity}

\begin{figure}[t]
	\centering
	\includegraphics[width=0.495\textwidth]{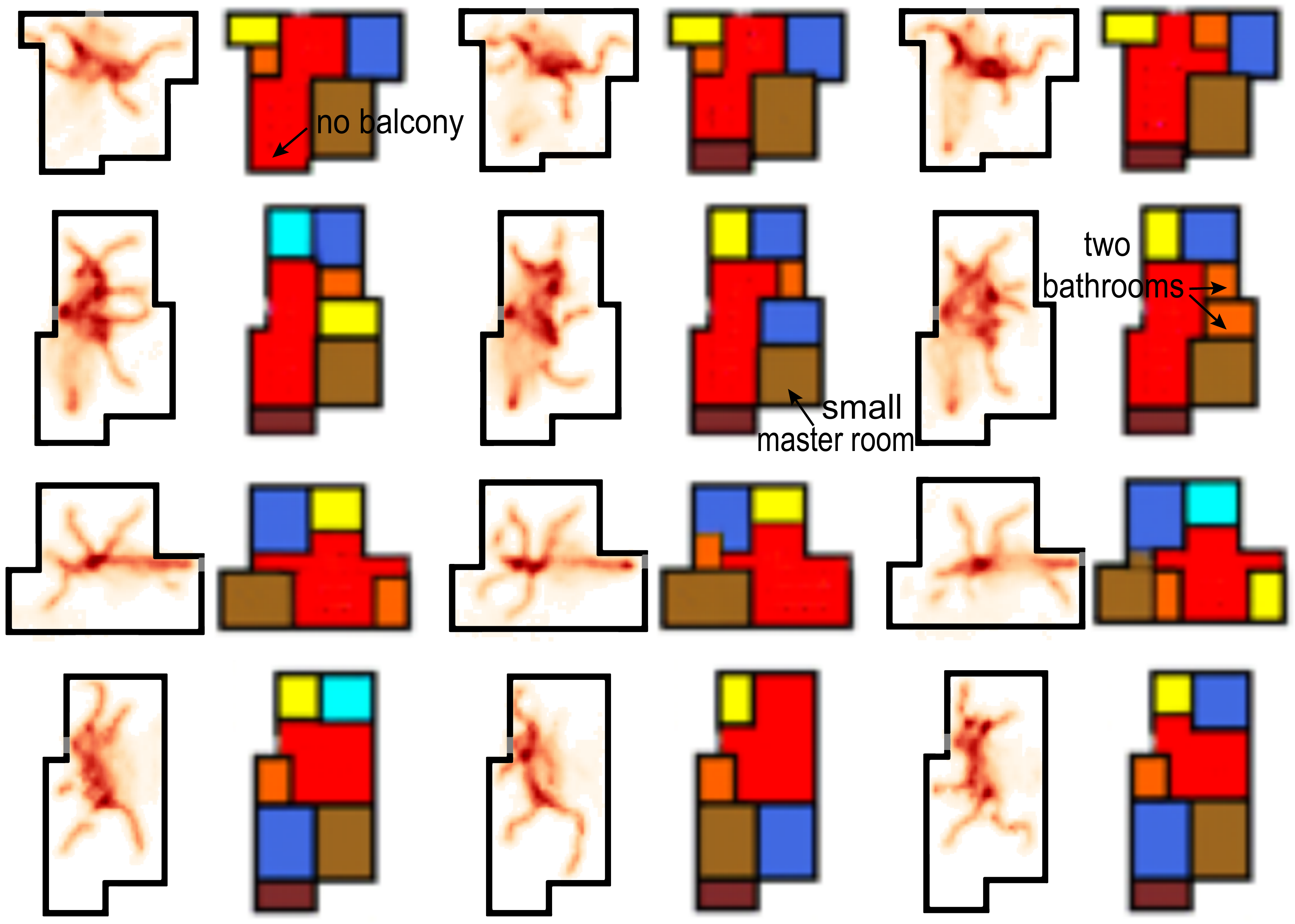} \\
	\caption{\tvcg{Our approach can generate diverse floorplans in guidance by different human-activity maps created by the stochastic generator with dropouts.}}
	\label{fig:diversity_GAN}
\end{figure}

\begin{figure*}[ht]
	\centering
	\includegraphics[width=0.995\textwidth]{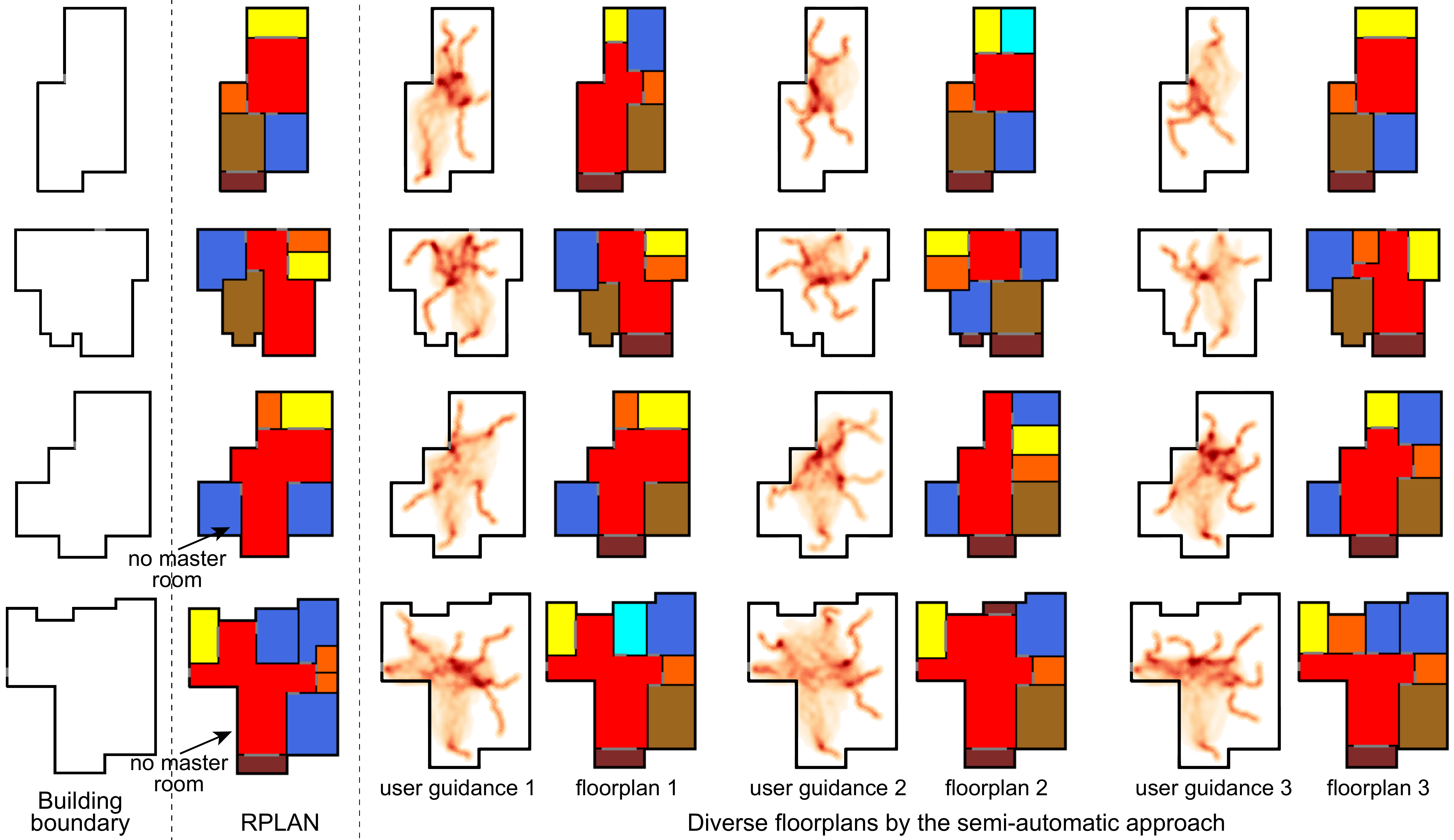} \\
	\caption{
		Our approach can generate diverse and high-quality human-centric floorplans for the same input building boundary, thanks to the semi-automatic approach that can produce different human-activity maps based on user demands.
	}
	\label{fig:diversity}
	\vspace*{-3mm}
\end{figure*}

\begin{figure}[t]
	\centering
	\includegraphics[width=8.35cm]{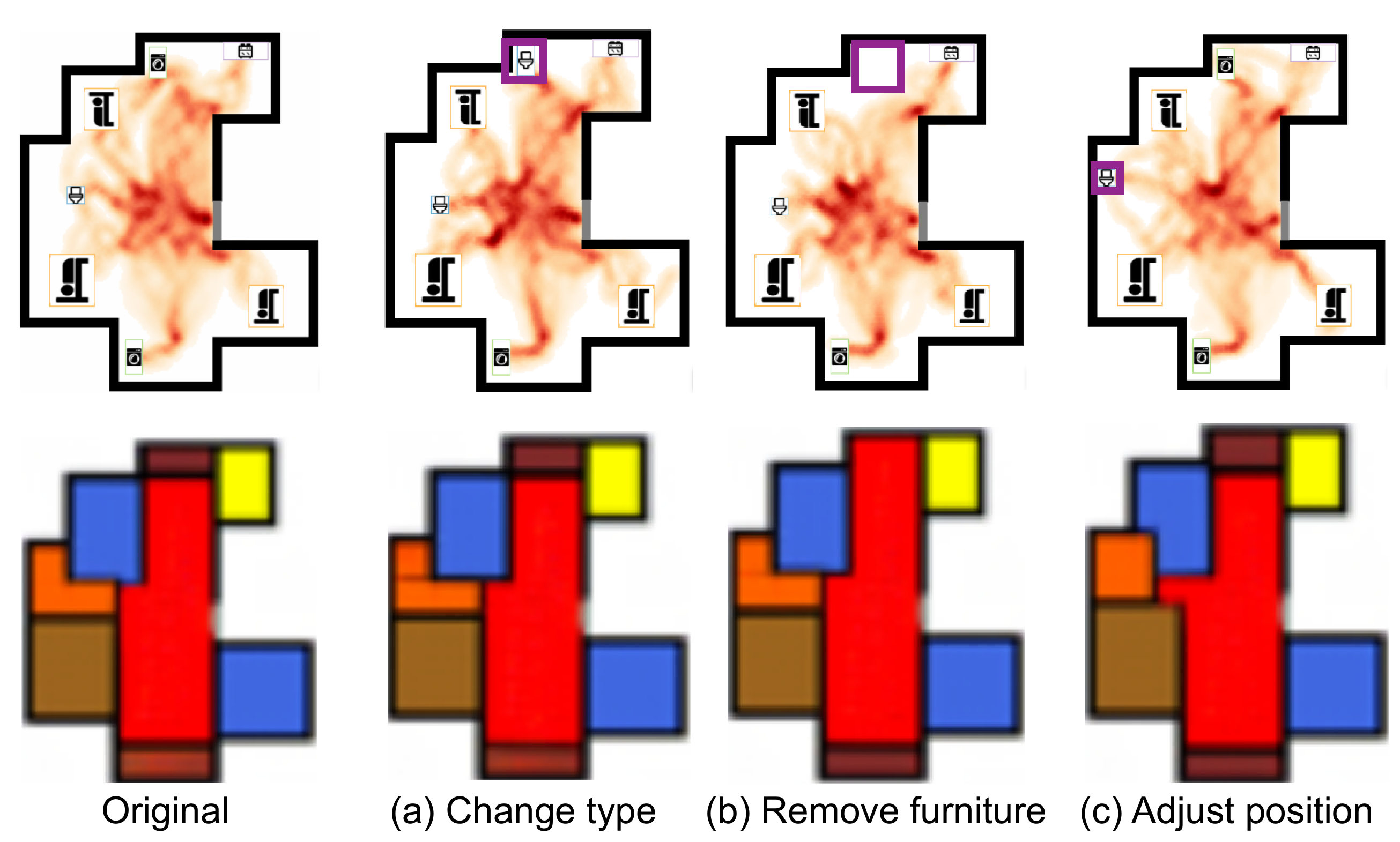} \\
	\vspace*{-2mm}
	\caption{\tvcg{Effects of different interactions applied to the furniture on human-activity maps and floorplans.}}
	\label{fig:interactive}
	\vspace*{-4mm}
\end{figure}

\tvcg{Our approach can generate diverse floorplans by using different human-activity maps as guidance, which are prepared either by the automatic approach with a generative network or by the semi-automatic approach with an interactive interface.}

\tvcg{Figure~\ref{fig:diversity_GAN} presents diverse floorplans produced by the automatic approach.
The generative network is able to produce stochastic human-activity maps using dropouts added to the first three layers of the decoder.
Here, the building boundaries are added as a reference.
Note that the generated human-activity maps are subtly different, yet the floorplans are rather diverse with different room numbers and types.
However, the floorplan designs may not be ideal, such as bad designs without balconies, a small master room, and two bathrooms next to each other.
The semi-automatic approach can overcome the deficiency by allowing users to manipulate the human-activity maps on demand.}

Figure~\ref{fig:diversity} presents diverse floorplans that can be derived from a single building boundary using the interactive interface (see Figure~\ref{fig:interface}).
For the same building boundary, the generated floorplans possess different properties, including the number of rooms, room positions and sizes, room types, and room adjacency relations, and they are different from the ground truths from RPLAN.
The human-activity maps are generated through our user interface by a user with no architecture background.
Not just being diverse, our results may also be more reasonable than real-world floorplans.
As shown in the last two rows of Figure~\ref{fig:diversity}, the floorplans from the RPLAN dataset have no master room but several second rooms, which are unnatural for common residential buildings.
Floorplans produced by our approach, instead, all have a master room. 

\tvcg{Figure~\ref{fig:interactive} illustrates the effects of applying different interactions to the furniture on the human-activity maps and floorplans.
In Figure~\ref{fig:interactive}(a), the washing machine is changed to a toilet.
Similar human-activity maps are generated like the original ones, and the floorplans have few changes.
In Figure~\ref{fig:interactive}(b), the washing machine is removed, leading to minimal activities in the area.
In this case, the balcony area in the original floorplan is merged with the living room.
In Figure~\ref{fig:interactive}(c), the toilet position is adjusted, and the corresponding shape and size of the toilet and the neighboring rooms are adjusted.
The results suggest that the human-activity map carries spatial information for room geometry guidance but no semantic information for inferring the room types.
}

The alternative approach of Wu et al.~\cite{wu_2019_data-driven} by default can only generate one single floorplan for an input building boundary.
To support the generation of diverse floorplans, they incorporate a probability distribution of predicted rooms, which, however, may not produce optimal floorplans.
Similar to our approach, Graph2Plan~\cite{hu_2020_Graph2Plan} provides a user interface that allows users to query and select a room topology graph, which supports the generation of diverse floorplans.
However, the retrieved graph needs to be adjusted to fit the building boundary, which requires domain knowledge to generate fine floorplans.
Nevertheless, even satisfactory room topology may not sufficiently capture the human spatial behavior and experience~\cite{franz_2008_from}.

\subsection{Discussion}
\label{ssec:discuss}
\begin{figure}[t]
	\centering
	\includegraphics[width=0.495\textwidth]{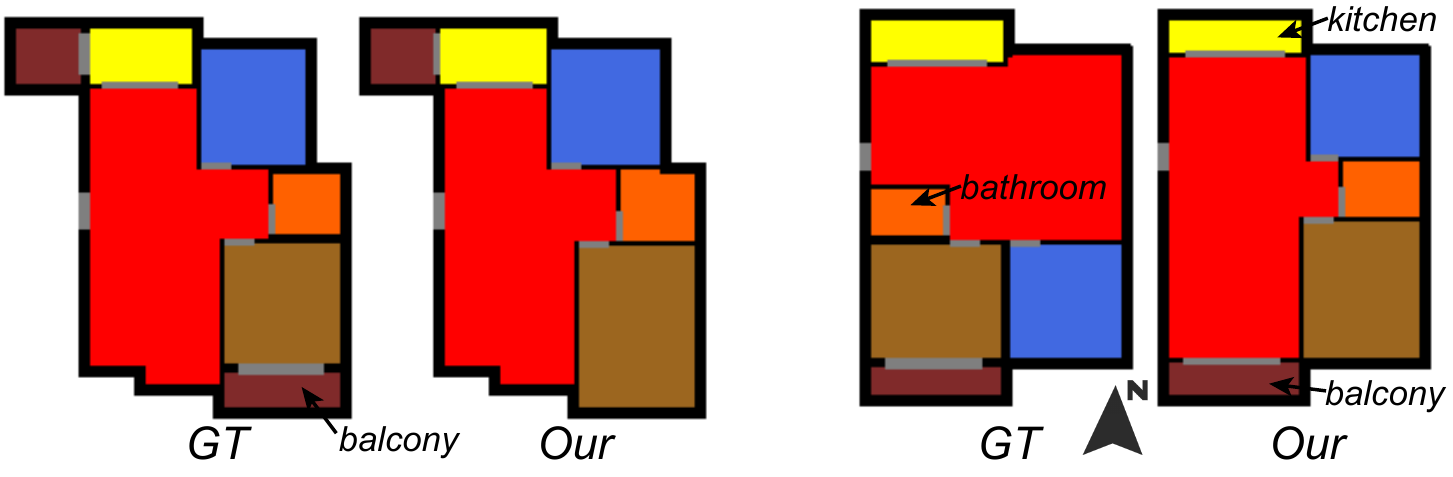} \\
	\vspace*{-2mm}
	\caption{Examples of \emph{GT} and \emph{Our} floorplans presented in the user evaluation by architects. \emph{Our} floorplan receives a lower rating on the left, and a higher rating on the right than \emph{GT}. }
	\label{fig:com_gt}
	\vspace*{-3mm}
\end{figure}

\emph{Quality}.
Figure~\ref{fig:com_gt} presents examples of \emph{GT} and our floorplans presented in the evaluation with architects presented in Sec.~\ref{ssec:quantitative}.
In Figure~\ref{fig:com_gt}(left), our floorplan receives a lower rating than \emph{GT}, though the layouts are almost the same.
The only difference is that our floorplan lacks an extra balcony outside the master room.
Even though the master room produced by our method is bigger, the interviewed architects explained that a master room with a balcony is typically more welcomed in practice.
In Figure~\ref{fig:com_gt}(right), our floorplan receives a higher rating than \emph{GT}.
In our floorplan, the balcony and kitchen are directly connected through a living room, whilst those in \emph{GT} are obstructed by a bathroom and a master room.
Our floorplan is a transparent layout that wind can easily blow through from south to north and vice versa, which is much preferred by the architects.

\vspace{1mm}
\noindent
\emph{Efficiency}.
Runtime is another important criterion for floorplan generation, as long delays can cause a bad user experience.
Our approach takes less than 50ms to predict and generate a piecewise floorplan from an input boundary, including around 30ms on generating the human-activity map using the automatic approach, and around 16ms for ActFloor-GAN module.
In contrast, Graph2Plan~\cite{hu_2020_Graph2Plan} takes about 178ms and Wu et al.~\cite{wu_2019_data-driven} require a total of about 4s for the network to process the input and generate a result.
Compared with the existing methods, our approach has an advantage in terms of time efficiency.

\begin{figure}[!t]
	\centering
	\includegraphics[width=0.495\textwidth]{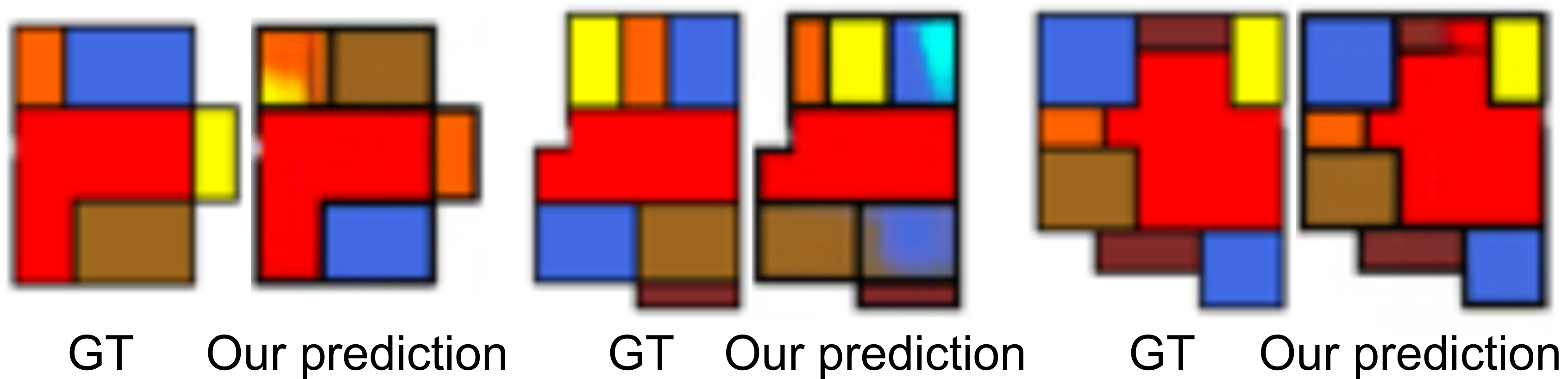} \\
	\caption{Examples of noisy ActFloor-GAN predictions.}
	\label{fig:failure}
	\vspace*{-3mm}
\end{figure}

\vspace{2mm}
\noindent
\emph{Limitations}.
Figure~\ref{fig:failure} shows some noisy predictions generated by ActFloor-GAN.
The left-hand side shows noisy kitchen and wall predictions for pixels in the upper-left corner, which should be a bathroom as shown in the \emph{GT} floorplan.
We notice that kitchen and bathroom pixels may be mistakenly predicted by ActFloor-GAN.
This is probably because kitchens and bathrooms in the RPLAN dataset are similar in terms of room size and position.
In the middle, there are noisy master room, second room, and study room predictions.
These three room types can be collectively referred to as bedroom, and it is rather difficult for a network to learn the slight difference between them. 
In fact, Graph2Plan~\cite{hu_2020_Graph2Plan} treats all these three room types as bedroom. 
However, floorplan design in practice needs to separate these room types, since the placement of balcony and bathroom often depends on the position of the master room.
On the right, predictions for the upper balcony are distracted by noisy predictions of the living room.
This is probably because there are two balconies directly connected to the living room as shown in the \emph{GT} floorplan, which is not common in the RPLAN dataset.
The network model is not able to catch such abnormal cases.

The human-activity map \TVCG{reflects room function and improves the prediction accuracy than vanilla models such as \emph{Pix2Pix} and \emph{CycleGAN}.
The more accurate prediction on pixel-wise room types also eases the vectorization process.}
\major{Besides, the activity map reflects the intensity of human activities in the rooms.
We find that the pixel of highest activity density in each room is highly correlated with position of the door, and use the information to place interior doors (see Figure~\ref{fig:post-processing}(c)).
In comparison, the current model of Graph2Plan~\cite{hu_2020_Graph2Plan} using room topology is not able to predict doors.}
Yet, further information, \TVCG{such as visibility and accessibility~\cite{berseth_2019_interactive}, which better indicate room function,} may help the network make noise-free predictions.

Another limitation of this work is the lack of model interpretability.
In Figure~\ref{fig:interactive}, we have shown that changes made to human-activity maps can lead to diverse floorplans, \major{and Figure~\ref{fig:compare_map} shows that the quality of final floorplans is highly affected by the synthesized human-activity maps}.
However, variations of the floorplans upon changes made to the human-activity maps are not always predictable.
Operations like removing furniture make big changes to human-activity maps, and ActFloor-GAN can predict a floorplan without the room.
Yet, small changes made by moving furniture or changing furniture types may not always lead to floorplans on desire.
Nevertheless, this is regarded as a general problem for existing deep learning models. 
Many efforts have been devoted to this direction.
We would like to incorporate emerging techniques to improve the interpretability of ActFloor-GAN in the near future.

\section{Conclusion and Future Work}
\label{sec:conclusion}
We presented ActFloor-GAN, a new deep framework for automated floorplan design.
Unlike existing deep-learning-based approaches that try to directly learn the geometric or topological properties of floorplans, we propose to tackle the problem from a new perspective, by leveraging the human-activity map as guidance for network training.
The benefit of introducing the human-activity map is prominent.
First, to incorporate the human-activity map in guiding the learning process, we re-formulate the cycle-consistent constraints in the generative networks.
We train ActFloor-GAN on the real-world floorplan dataset RPLAN, and conduct experiments showing that the guidance by the human-activity map and the cycle-consistent constraints can produce more accurate network predictions.
Second, the human-activity map reflects resident interaction with the environment, so the generated floorplans by our approach are both geometrically plausible and topologically reasonable.
As evaluated by five architects, the generated floorplans satisfy the requirement of being human-centric, and of the compelling quality with the professionally-designed ones.
Third, the human-activity map depends on the room layout and furniture locations, so different human-activity maps can be derived by manipulating the furniture placement using a user interface.
Correspondingly, we can generate diverse floorplans for the same given building boundary.

There are several promising directions for future work.
Though the human-activity map improves the overall performance, its contributions to network prediction vary upon the room type.
As shown in the failure cases, ActFloor-GAN may mistakenly mix kitchen and bathroom, and also the master room, second room, and study room. 
This is probably because our representation of the human-activity map is essentially a probability map of human spatial behavior.
The collaborating architect suggests utilizing other architectural metrics that can encode further details on the human-environment interactions, which have been extensively studied in the space syntax~\cite{hillier_1976_space}.
We envision more accurate network predictions with these additional metrics.
Moreover, we would like to extend the current framework from residential buildings to other scenarios like office space and public space design, which shall consider substantially more people movements in the planning stage~\cite{hillier_1993_natural}.

\section*{Acknowledgments}
The authors wish to thank the anonymous reviewers for their valuable comments.
This work is supported in part by the Fundamental Research Funds for the Central Universities (22120210540) and the Research Grants Council of the Hong Kong Special Administrative Region (Project no. CUHK 14206320).
	
\bibliographystyle{abbrv}
\bibliography{Reference}

\begin{IEEEbiography}[{\includegraphics[width=1in,height=1.25in,clip,keepaspectratio]{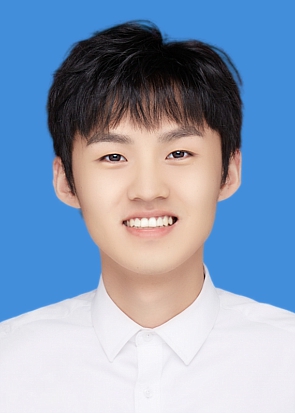}}]
{Shidong Wang} was a master student at the School of Computer Science and Technology, Shandong University, and was a visiting student at Shenzhen Institute of Advanced Technology, Chinese Academy of Sciences.
His research interests include human-centric AI and computational design.
\end{IEEEbiography}

\begin{IEEEbiography}[{\includegraphics[width=1in,height=1.25in,clip,keepaspectratio]{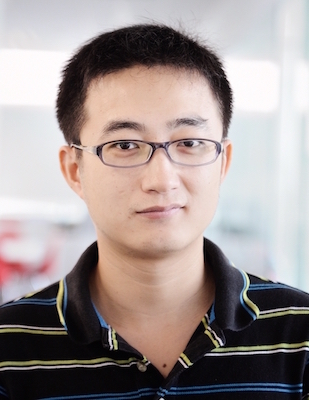}}]
{Wei Zeng} is an assistant professor in the Hong Kong University of Science and Technology, Guangzhou campus.
He received the PhD degree in computer science from Nanyang Technological University, and worked as a senior research at Future Cities Laboratory, ETH Zurich, and an associate researcher at Shenzhen Institute of Advanced Technology, Chinese Academy of Sciences.
His recent research interests include visualization and visual analytics, computer graphics, AR/VR, and HCI.
\end{IEEEbiography}

\begin{IEEEbiography}[{\includegraphics[width=1.0in,height=1.25in,clip,keepaspectratio]{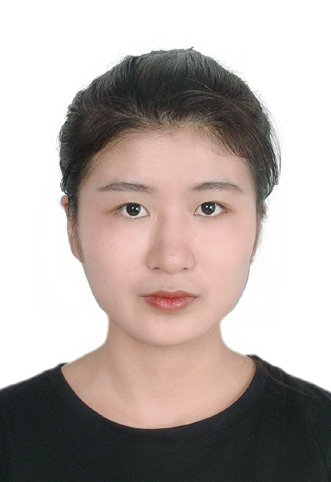}}]
{Xi Chen} is a master student at Shenzhen Institute of Advanced Technology, Chinese Academy of Sciences, and also with University of Chinese Academy of Sciences.
Her research interests include data visualization and human computer interaction.
\end{IEEEbiography}

\begin{IEEEbiography}[{\includegraphics[width=1.0in,height=1.25in,clip,keepaspectratio]{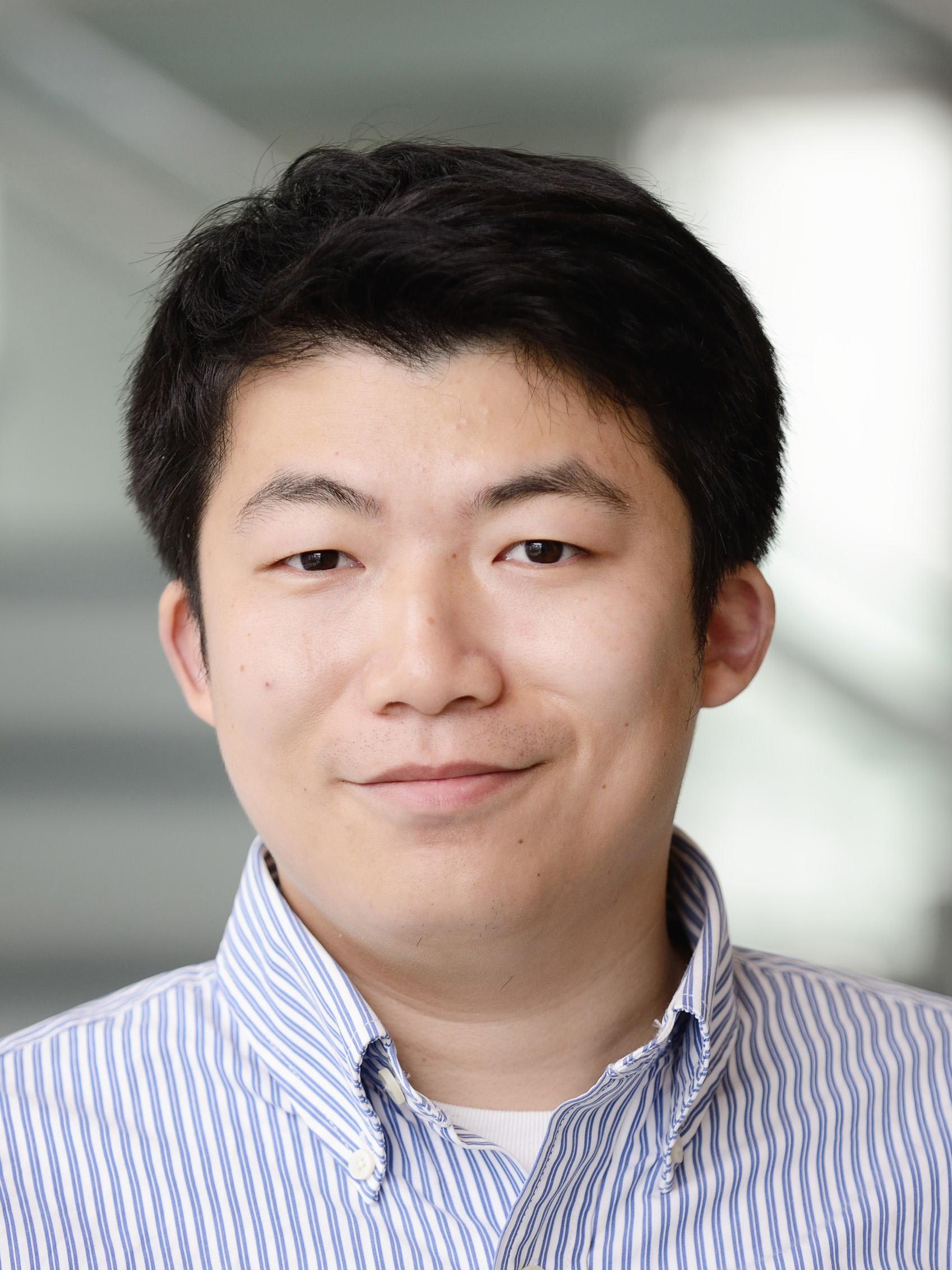}}]{Yu Ye} is
an associate professor in the Department of Architecture, College of Architecture and Urban Planning at Tongji University, China. He gained his PhD degree from The University of Hong Kong and he worked as a Post-doctoral researcher at Future Cities Laboratory, ETH Zurich. His research mainly focuses on computational urban design utilizing multi-sourced urban data and newly emerged analytical techniques.
\end{IEEEbiography}

\begin{IEEEbiography}[{\includegraphics[width=1.0in,height=1.25in,clip,keepaspectratio]{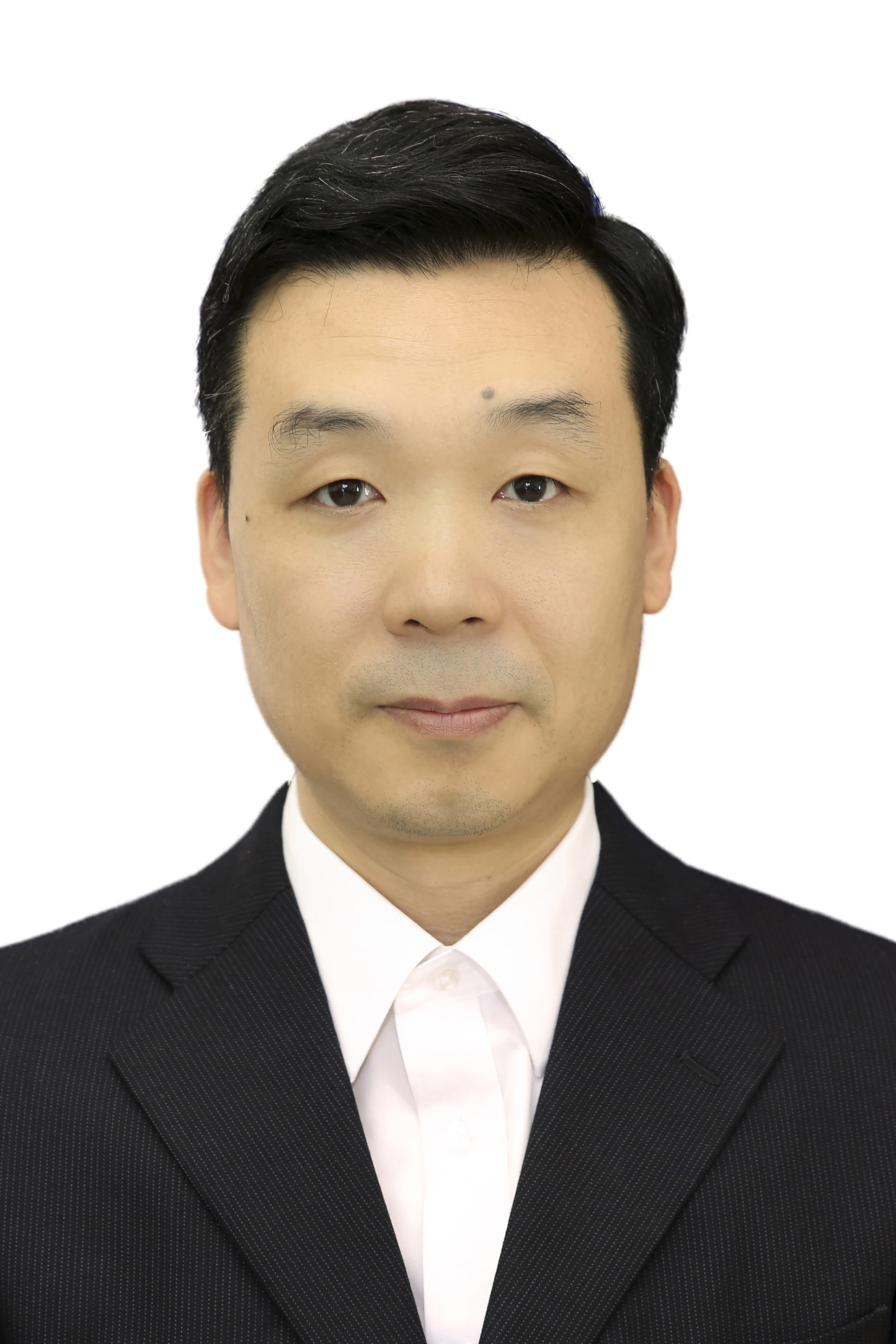}}]
{Yu Qiao} is a professor with Shenzhen Institute of Advanced Technology, Chinese Academy of Science, and the director of Institute of Advanced Computing and Digital Engineering. His research interests include computer vision, deep learning, and bioinformation. He received the first prize of Guangdong technological invention award, and Jiaxi Lv young researcher award from Chinese Academy of Sciences. He served as the program chair of IEEE ICIST 2014.\end{IEEEbiography}

\begin{IEEEbiography}[{\includegraphics[width=1.0in,height=1.25in,clip,keepaspectratio]{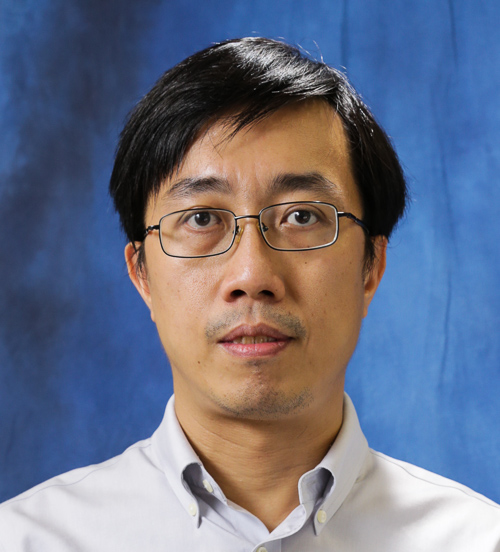}}]{Chi-Wing Fu} is a professor in the Chinese University of Hong Kong. He served as the co-chair of SIGGRAPH ASIA 2016's Technical Brief and Poster program, associate editor of IEEE Computer Graphics \& Applications and Computer Graphics Forum, panel member in SIGGRAPH 2019 Doctoral Consortium, and program committee members in various research conferences, including SIGGRAPH Asia Technical Brief, SIGGRAPH Asia Emerging tech., IEEE visualization, CVPR, IEEE VR, VRST, Pacific Graphics, GMP, etc.  His recent research interests include computation fabrication, point cloud processing, 3D computer vision, user interaction, and data visualization.
\end{IEEEbiography}

\vfill

\end{document}